\documentclass{article}
\pdfoutput=1 

\usepackage{arxiv}

\usepackage[utf8]{inputenc} 
\usepackage[T1]{fontenc}    
\usepackage{hyperref}       
\usepackage{url}            
\usepackage{booktabs}       
\usepackage{amsfonts}       
\usepackage{nicefrac}       
\usepackage{microtype}      
\usepackage{lipsum}
\usepackage{graphicx}
\usepackage{algpseudocode}
\usepackage{bmpsize}

\usepackage{bm,amsmath,bbm,amsfonts,amsthm,amsbsy,amscd,amsxtra,amsgen,amsopn,amssymb,mathtools,caption, subcaption}
\usepackage{threeparttable}
\usepackage[usenames,dvipsnames]{xcolor}
\usepackage{natbib}

\newcommand{\bs}[1]{\boldsymbol{#1}}
\newtheorem{theorem}{Theorem}

\newtheorem{algorithm}{Algorithm}

\title{Adaptive enrichment trial designs using joint modeling of longitudinal and time-to-event data}

\author{Abigail J. Burdon$^{1,*}$, 
Richard D. Baird$^{2}$, and Thomas Jaki$^{1,3}$ \\
$^{1}$MRC Biostatistics Unit, University of Cambridge, Robinson Way, Cambridge, CB2 0SR, U.K\\
$^{2}$Cancer Research UK, Cambridge Centre, University of Cambridge, Robinson Way, Cambridge, CB2 0RE, U.K\\
$^{3}$University of Regensburg, Bajuwarenstrasse 4, 93053 Regensburg, Germany \\
$^{*}$\texttt{Email: abigail.burdon@mrc-bsu.cam.ac.uk}}

\begin{document}
\maketitle
\begin{abstract}
Adaptive enrichment allows for pre-defined patient subgroups of interest to be investigated throughout the course of a clinical trial. Many trials which measure a long-term time-to-event endpoint often also routinely collect repeated measures on biomarkers. In this work, we present a joint model for longitudinal and time-to-event data and two methods for creating standardised statistics based on this joint model. We then use the estimates to define subgroup selection rules and efficacy and futility early stopping rules for a flexible efficient clinical trial with possible enrichment and show that the familywise error rate is protected in the strong sense. To assess the results, we consider a trial for the treatment of metastatic breast cancer where repeated circulating tumour DNA (ctDNA) measurements are available and the subgroup criteria is defined by patients' HER2 status. We show that incorporating biomarker information leads to accurate subgroup identification and increases in power.
\end{abstract}


\maketitle

\section{Introduction}
In current oncology practice and cancer clinical trials, it is crucial to focus testing of novel therapies on the patient subgroups most likely to benefit. Too many patients receive treatments that either do not work particularly well, are toxic, or sometimes both.  Adaptive enrichment clinical trials enable the efficient testing of an experimental intervention on specific patient subgroups of interest (see~\citep{burnett2020adding} and~\citep{pallmann2018adaptive}). At an interim analysis, if a particular subgroup of patients is identified as responding particularly well to treatment, then we can focus resources and inferences by recruiting additional patients from the subgroup which benefits. \citet{simon2013adaptive} show the benefits of enrichment trials, in particular that patients who do not appear to benefit are removed from the experimental treatment with potentially harmful side effects. If the treatment is futile for all patients, we are able to terminate the trial at interim analyses~\citep{burnett2021adaptive}. Further, if patients respond overwhelmingly well to treatment, then there is potential to stop the trial early for efficacy demonstrating that the experimental treatment is superior to control in this subgroup, and the usual benefits of group sequential tests apply~\citep{jennison2000group}.

In recent years, there has been increased uptake in enrichment trials which consider a long-term time-to-event (TTE) endpoint, such as overall survival (OS), but this is still low compared to continuous endpoints~\citep{ondra2016methods}. In such trials, it is common for investigators to also collect repeated measures on biomarkers. Recent research proposes methods which use the short-term endpoint data for subgroup selection rules then focus on the primary endpoint data for hypothesis testing~\citep{stallard2010confirmatory,friede2011designing}. Our aim is to leverage the additional biomarker information to improve interim decision making, early stopping rules \emph{and} hypothesis testing.

We present a joint model for longitudinal and TTE data and base an enrichment trial design on the treatment effect in the joint model. There has been significant interest in joint modeling of longitudinal and time-to-event data~\citep{henderson2000joint,rizopoulos2012joint} with a focus on prediction and personalised medicine. However, the uses of joint modeling have yet to be established in clinical trial designs. We show that by incorporating the longitudinal data into the analysis via joint modeling, this results in the subgroup which benefits being selected more frequently and higher power (using the same number of patients) as the equivalent trial which ignores the biomarker observations. Our simulation results are based on data from a study which measured OS and plasma circulating tumour DNA (ctDNA) levels ~\citep{dawson2013analysis}. To define subgroups, we hypothesise that patients who are HER2 negative will benefit from the experimental treatment more than patients who are HER2 positive.

Similarly to \citet{magnusson2013group}, we use the ``threshold selection" rule combined with an error spending test to clearly predefine the subgroup selection and stopping rules before the trial commences. We present a method where, in the setting of TTE data and joint modeling, the relationship between number of observed events and information levels can be exploited to design an efficient clinical trial. The novel feature of this work is an enrichment trial which uses a modern joint model to make both interim decisions and perform hypothesis testing.
\section{Motivating example}
\label{sec:motivation}
Fragments of ctDNA are detected in the blood of cancer patients and are routinely measured in many cancer clinical trials. These measurements, which we shall often refer to as ``biomarker measurements" or ``longitudinal data" are useful prognostic factors that can improve the precision of OS estimates. Throughout this paper, we shall base our analyses on data from a study which compared different biomarkers and their accuracy in monitoring tumour burden among women with metastatic breast cancer~\citep{dawson2013analysis}. The results of the study were conclusive that ctDNA was successfully detected and highly correlated with OS.

Another important factor in breast cancer studies is the presence or absence of the HER2 protein. Patients who are HER2 positive may be resistant to conventional therapies and treatments that specifically target the HER2 protein are very effective~\citep{slamon1987human}. Adaptive enrichment trials are therefore highly efficient in breast cancer settings because the eligibility criteria based on HER2 status can be updated during the trial, restricting entry to patients likely to benefit.

\section{Joint modeling of ctDNA and OS in defined subgroups}
\label{sec:joint}
\subsection{Subgroup set-up and notation}
For adaptive enrichment trials, a key assumption is that subgroup identification is known prior to commencement. For the metastatic breast cancer example of Section~\ref{sec:motivation}, let $S_1$ denote the HER2 negative subgroup and let $S_2$ denote the HER2 positive subgroup. Then, let $F=S_1\cup S_2$ denote the full population. Extensions to more subgroups can be made following the same logic. Further, we denote $K$ as the total number of analyses in the adaptive trial and for our metastatic breast cancer example, we shall use $K=2.$ 

The aim of a clinical trial is to assess how effective a new experimental treatment performs compared to an existing standard-of-care drug or placebo. We make statistical
inferences based on a treatment effect $\theta$ which is defined at the design stage. For a trial with multiple subgroups, let $\theta_j$ be the treatment effect in subgroup $j=1,2,F$. A mathematical consequence is that if the prevalence of $S_1$ in $F$ is given by $\lambda$, then 
\begin{equation}
\label{eq:theta_full}
\theta_F = \lambda\theta_1+(1-\lambda)\theta_2.
\end{equation}
Throughout, it is assumed that $\lambda$ is known and fixed, however methods are available that account for uncertainty and allow estimation of $\lambda$ at each analysis~\citep{wan2019subgroup}. We aim to test the hypotheses
\begin{equation}
\label{eq:hypothesis}
H_{0,j}: \theta_j \leq 0\; \text{ vs }\; H_{A,j}:\theta_j > 0 \hspace{1cm} \text{for } j=1,2,F.
\end{equation}

\subsection{The joint model}
\label{subsec:joint_model}
The joint model that we consider is an adaptation of Equation (2) of \citet{tsiatis2001semiparametric} (referred to as ``TD" for short). There are two processes in this model which represent the survival and longitudinal parts separately, and these processes are linked using random effects. Let the times of the measurements of the longitudinal data for patient $i$ in subgroup $j=1,2$ be denoted by $v_{ji1},\dots v_{jim_{ji}}$, then $X_{ji}(v_{jis})$ is the true value of the biomarker at time $v_{jis}$ and $W_{ji}(v_{jis})$ is the observed value of the biomarker. The longitudinal model takes the form
\begin{equation}
\begin{split}
\label{eq:long1}
X_{ji}(v_{jis})&=b_{0ji}+b_{1ji}v_{jis} \\
W_{ji}(v_{jis})&= X_{ji}(v_{jis})+\epsilon_{ji}(v_{jis})
\end{split}
\end{equation}
where $\mathbf{b}_{ji}=(b_{0ji},b_{1ji})$ is a vector of patient specific random effects and $\epsilon_{ji}(v_{jis})$ is the measurement error. We make the assumptions that $\epsilon_{ji}(v_{jis})|\mathbf{b}_{ji}\sim N(0,\sigma_j^2) \text{ for } s=1,\dots,m_{ji}$ and $\epsilon_{ji}(v)$ and $\epsilon_{ji}(v')$ are independent for $v\neq v'.$ We have chosen to model the longitudinal data as linear in time. This appears appropriate for the example dataset of Section~\ref{sec:motivation} as we have seen ctDNA display this property. The methods can easily be extended to incorporate more complex trajectories for the longitudinal data.

The model for the survival endpoint is a Cox proportional hazards model. Suppose that $\psi_{ji}$ is the indicator function that patient $i$ in subgroup $j=1,2$ receives the experimental treatment. Let $\theta_j$ and $\gamma_j$ be a scalar coefficients. Then the hazard function for subgroup $j$ is given by
\begin{equation}
\label{eq:surv}
h_{ji}(t)=h_{0j}(t)\exp\{\gamma_j X_{ji}(t)+\theta_j \psi_{ji}\} \hspace{1cm} \text{for }j=1,2.
\end{equation}
Note that it is the biomarker value $X_{ji}(v_{jis})$ which is included as a variable in the proportional hazards model, whereas the measurements $W_{ji}(v_{jis})$ with added error are observed. Together, Equations~\eqref{eq:long1}--\eqref{eq:surv} define the joint model and defines the working model from which we shall perform simulation studies in Section~\ref{sec:results}. We note here that there is no treatment effect included in the biomarker trajectory. The motivation for this follows the models that are presented in the literature given by TD. For a more general model including a treatment effect in the longitudinal data, we refer the reader to the supplementary materials where we discuss the use of the Restricted Mean Survival Time (RMST) endpoint which can account for multiple treatment effect parameters. The RMST methodology requires additional modelling assumptions and  performs poorly under model misspecification, and for this reason we do not consider it further in the main text.

\subsection{Conditional score}
\label{subsec:cond_score}
To perform the adaptive enrichment trial, we must find treatment effect estimates and their distributions at analyses $k=1,\dots,K$ and subgroups $j=1,2,F.$ To do so, we shall use a modified version of the conditional score method by TD which is a method for fitting the joint model to the data. We present multi-stage adaptations of some functions presented in TD. Let $t_{ji}^{(k)}$ be the observed event time and let $\delta_{ji}^{(k)}$ be the observed censoring indicator for patient $i$ in subgroup $j=1,2$ at analysis $k$. This censoring event includes censoring patients who remain in the study at analysis $k$ but have not yet observed the event at the given analysis. We denote the maximum follow-up time at analysis $k$ by $\tau_k$. To be included in the at-risk set at time $t$, the patient must have at least two longitudinal observations to fit the regression model. At analysis $k$, we define the at-risk process, $Y_{ji}^{(k)}(t)=\mathbb{I}\{t_{ji}^{(k)}\geq t, v_{ji2}\leq t\} $, counting process, $N_{ji}^{(k)}(t)=\mathbb{I}\{t_{ji}^{(k)} \leq t, \delta_{ji}^{(k)} = 1, v_{ji2} \leq t\}$ and function $dN_{ji}^{(k)}(t)= \mathbb{I}\{t\leq t_{ji}^{(k)} < t+dt, \delta_{ji}^{(k)} = 1, v_{ji2} \leq t\}$ for the joint model.

An object of importance is the sufficient statistic. For patient $i$ in subgroup $j$, let $\hat{X}_{ji}(v)$ be the ordinary least squares estimate of $X_{ji}(v)$ based on the set of measurements taken at times $\{v_{ji1},\dots,v_{jis}|v_{jis}\leq v\}$ and suppose that $\sigma_j^2\psi_{ji}(v)$ is the variance of $\hat{X}_{ji}(v)$. TD define the sufficient statistic to be the function
$$
S_{ji}^{(k)}(t,\gamma_j,\sigma_j^2)=\hat{X}_{ji}(t) + \gamma_j\sigma_j^2\psi_{ji}(t)dN_{ji}^{(k)}(t)
$$
which is defined for all $t\in (v_{ji2},t_{ji}^{(k)})$ for patient $i$ in subgroup $j$. The multi-stage version of the scalar $E_{0i}$ of TD, dependent on subgroup $j$, is given by 
$$
E_{0ji}^{(k)}(t, \gamma_j,\theta_j, \sigma_j^2) = \exp\{\gamma_j S_{ji}^{(k)}(t,\gamma_j,\sigma_j^2)-\gamma_j^2\sigma_j^2\psi_{ji}(t)/2+\theta_j \psi_{ji}\}
$$
and the multi-stage version of the quotient function $E_1/E_0$ in Equation~(6) by TD, dependent on subgroup $j$, is the $2\times1$ vector given by
$$
E_j^{(k)}(t, \gamma_j,\theta_j, \sigma_j^2) =
\frac{\sum_{i=1}^{n_j} \{S_{ji}^{(k)}(t, \gamma_j,\sigma_j^2), \psi_{ji}\}^T E_{0ji}^{(k)}(t, \gamma_j,\theta_j, \sigma_j^2) Y_{ji}^{(k)}(t)}{
\sum_{i=1}^{n_j}E_{0ji}^{(k)}(t, \gamma_j,\theta_j, \sigma_j^2)Y_{ji}^{(k)}(t)}.
$$
Then, the conditional score function at analysis $k$ for subgroup $j=1,2$, also a vector of dimension $2\times 1$, is given by
\begin{equation}
\begin{split}
\label{eq:score_1}
U_j^{(k)}&(\gamma_j,\theta_j,\sigma_j^2) = \\
&\int_0^{\tau_k}\sum_{i=1}^{n_j}\left( \{S_{ji}^{(k)}(t, \gamma_j,\sigma_j^2), \psi_{ji}\}^T
- E_j^{(k)}(t, \gamma_j,\theta_j, \sigma_j^2)\right) dN_{ji}^{(k)}(t).
\end{split}
\end{equation}

\subsection{Estimates for the treatment effects $\theta_j$ and their distributions}

The aim is now to find treatment effect estimates $\hat{\theta}^{(k)}_j$ for $j=1,2,F$ and analyses $k=1,\dots,K.$ We define these estimates as the root of the conditional score. In doing so, it turns out that these estimates are asymptotically normally distributed and we derive the variance of the estimates.

\citet{https://doi.org/10.48550/arxiv.2211.16138} show that $\mathbb{E}(U_j^{(k)}(\gamma_j,\theta_j,\sigma_j^2))=\mathbf{0}$  for each $k=1,\dots,K,$ and $j=1,2$. Therefore, the conditional score function at analysis $k$ and subgroup $j=1,2$ is an estimating function, and set equal to zero defines an estimating equation. Hence, asymptotically normal parameter estimates for $\gamma_j$ and $\theta_j$ can be found as the root of the estimating equation. As in TD Equation~(13), define the pooled estimate $\hat{\sigma}_j^{(k)2}=\sum_{i=1}^{n_j} \mathbb{I}\{m_{ji}(k)>2\}R_{ji}(k)/\sum_{i=1}^{n_j}\mathbb{I}\{m_{ji}(k)>2\}(m_{ji}(k)-2),$ where $R_{ji}(k)$ is the residual sum of squares for the least squares fit to all $m_{ji}(k)$ observations for patient $i$ in subgroup $j$ available at analysis $k$. Then, let $\hat{\gamma}_j^{(k)},\hat{\theta}_j^{(k)}$ be the values of $\gamma_j$ and $\theta_j$ respectively such that
$$
U_j^{(k)}(\hat{\gamma}_j^{(k)},\hat{\theta}_j^{(k)},\hat{\sigma}_j^{(k)2}) = \mathbf{0}.
$$

We also need to know the distribution of these estimates and this requires knowledge of the variance of $\hat{\theta}_j^{(k)}$. We shall use the sandwich estimator, as in Section 2.6 by \citet{wakefield2013bayesian}, to calculate a robust estimate for the variance of the parameter estimates. Firstly, define matrices
\begin{align*}
A_j^{(k)} &=\partial U_j^{(k)}(\gamma_j,\theta_j,\sigma_j^2)/\partial(\gamma_j,\theta_j)^T \\
B_j^{(k)} &= Var(U^{(k)}_j(\gamma_j,\theta_j,\sigma_j^2)).
\end{align*}

\citet{https://doi.org/10.48550/arxiv.2211.16138} present analytical forms for each of these $2\times 2$ matrices including a detailed calculation for the derivative matrix $A_j^{(k)}.$ In practice, $A_j^{(k)}$ can be calculated numerically and $B_j^{(k)}$ is found by considering the conditional score as a sum over $n_j$ patients. Further, these matrices are estimated by substituting the estimates $\hat{\gamma}_j^{(k)}, \hat{\theta}_j^{(k)}$ and $\hat{\sigma}_j^{(k)2}$ for $\gamma_j,\theta_j$ and $\sigma_j^2$ respectively. Then the information for the treatment effect estimate is given by
$$
\mathcal{I}_j^{(k)}=1/Var(\hat{\theta}^{(k)}_j) = n_j\left[ (A_j^{(k)})^{-1}B_j^{(k)}((A_j^{(k)})^{-1})^T\right]^{-1}_{22}
$$
for $j=1,2$ and $k=1,\dots,K.$ The subscript represents that we are interested in the second parameter $\theta_j$ in the vector $(\gamma_j,\theta_j,\sigma_j^2)^T.$

In accordance with Equation~\eqref{eq:theta_full}, the treatment effect estimate and corresonding information level in the full population at analysis $k=1,\dots,K$ are given by
\begin{align*}
\hat{\theta}^{(k)}_F &= \lambda\hat{\theta}^{(k)}_1+(1-\lambda)\hat{\theta}^{(k)}_2 \\
\mathcal{I}^{(k)}_F &= \left(\lambda^2/\mathcal{I}^{(k)}_1 + (1-\lambda)^2/\mathcal{I}^{(k)}_2\right)^{-1}.
\end{align*}
Finally, standardised $Z-$statistic are given by
$$
Z^{(k)}_j = \hat{\theta}^{(k)}_j\sqrt{\mathcal{I}^{(k)}_j} \hspace{1cm} \text{for }j=1,2,F \text{ and } k=1,\dots,K.
$$

For simplicity in notation and exposition, we now return to the example of Section~\ref{sec:motivation} in which $K=2.$ In order for subsequent results to hold, we require $Z^{(1)}_j,Z^{(2)}_j$ to have the ``canonical joint distribution" (CJD) given in Section 3.1 of \citet{jennison2000group} for each $j=1,2,F.$ The canonical joint distribution of the standardised statistics across analyses is such that
\begin{equation}
\label{eq:CJD}
\begin{bmatrix}Z^{(1)}_j \\ Z^{(2)}_j \end{bmatrix}\sim N\left( \begin{bmatrix} \theta^{(1)}_j\sqrt{\mathcal{I}^{(1)}_j} \\ \theta^{(2)}_j\sqrt{\mathcal{I}^{(2)}_j} \end{bmatrix}, \begin{bmatrix} 1 & \sqrt{\mathcal{I}^{(1)}_j/\mathcal{I}^{(2)}_j} \\   \sqrt{\mathcal{I}^{(1)}_j/\mathcal{I}^{(2)}_j} & 1\end{bmatrix} \right).
\end{equation}
\citet{https://doi.org/10.48550/arxiv.2211.16138} show that the $Z$-statistics calculated using the conditional score methodology have approximately the the canonical joint distribution, but not exactly. The authors show that by proceeding with a group sequential test assuming that this does hold is sensible since Type 1 error rates are conservative and diverge minimally from planned significance level. We give simulation evidence that this is also true for an adaptive enrichment trial in Section~\ref{sec:results}.

\section{Adaptive enrichment schemes for clinical trials with subgroup selection}
\label{sec:enrichment_theory}

\subsection{The threshold selection rule}
\label{subsec:threshold} 
\begin{figure}
\centering\includegraphics[width=\textwidth]{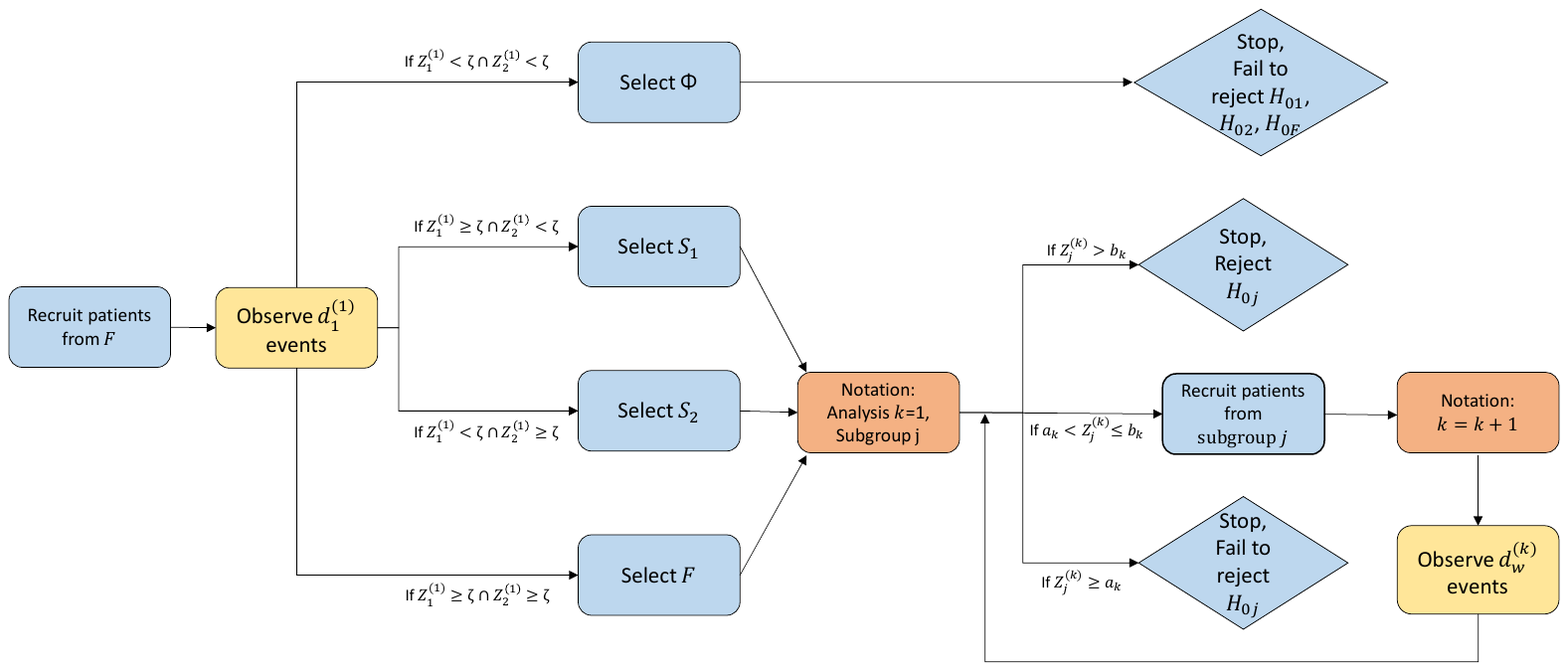}
\caption{Flow chart for enrichment trial design which uses the threshold rule for subgroup selection at the interim analysis. Hypothesis testing is based on an error spending design with $\alpha$-spending for the efficacy boundary and $\beta$-spending for the futility boundary including the opportunity for early stopping. The flow chart describes when the interim analysis should be performed based on the pre-planned number of events $d^{(1)}_1$ in subgroup $S_1$ at the interim and the total number of observed events $d^{(2)}$ in the selected subgroup at the final analysis.}
\label{fig:flow_chart}
\end{figure}

An adaptive enrichment scheme consists of two decisions; firstly a decision upon which subgroup, if any, to continue the trial with at the interim analysis and secondly, a decision upon whether or not to reject the null hypothesis at the final analysis. There are a collection of rules which can be used for subgroup selection, for example the maximum test statistic~\citep{chiu2018design} and a Bayes optimal rule~\citep{burnett2021adaptive}. Similarly to \citet{magnusson2013group}, we shall use the threshold selection rule. The definition is as follows; for some constant $\zeta$, select all subgroups $j\in\{1,2\}$ such that $Z_j^{(1)} > \zeta$ (Figure~\ref{fig:flow_chart}). If $Z_1^{(1)} > \zeta$ and $Z_2^{(1)} > \zeta$ then the trial continues in the full population and it should be noted that this is a stronger condition than $Z_F^{(1)} > \zeta$ as in the latter case, overwhelming benefit in one subgroup with poor effect in the other could still lead to selection of the full population. Finally, if $Z_1^{(1)} \leq \zeta$ and $Z_2^{(1)} \leq \zeta$ then the trial stops at the interim analysis declaring the treatment to be in-efficacious in all subgroups. This ensures that only subgroups which have a large enough treatment effect are followed to the second analysis. The threshold selection rule leads to an efficient enrichment trial design because we can find analytical forms for the Type 1 and Type 2 error rates and are therefore able to maximise power. As well as clearly providing the generic design framework for any test statistic, a novel aspect of this work will be applying this rule in the joint modeling setting.

For the choice of $\zeta$, we impose some restrictions which can be customised at the design stage of the trial. First, we define the configuration of parameters under the global null as $\bs{\Theta}_G:\{\theta_1=\theta_2=\theta_F=0\}$ and the alternative as $\bs{\Theta}_A:\{\theta_1=\delta,\theta_2=0,\theta_F=\lambda\delta\}$. This represents that we believe there is an important effect of treatment in $S_1$. For the metastatic breast cancer example in Section~\ref{sec:motivation}, this reflects that the HER2 negative subgroup is expected to respond well to the treatment. We require that, under $H_A$, it is equally likely to select the full population or no subgroup. We also require that the subgroup which truly responds well to treatment, is selected a high proportion of times, denoted $\psi$. Hence, we solve the simultaneous equations 
\begin{align*}
\mathbb{P}_{\bs{\Theta}_A}(Z^{(1)}_1 >\zeta \cap Z^{(1)}_2 \leq \zeta) &= \psi \\
\mathbb{P}_{\bs{\Theta}_A}(Z^{(1)}_1 >\zeta \cap Z^{(1)}_2 > \zeta) &=\mathbb{P}(Z^{(1)}_1 \leq \zeta \cap Z^{(1)}_2 \leq \zeta).
\end{align*}
As an example, with $\psi=0.6$ and $\delta=-0.5$ and using the distribution of the test statistics given by Equation~\eqref{eq:CJD}, we therefore need $\zeta=0.754$ and $\mathcal{I}^{(1)}_1=9.08$.

We now present the joint distribution of the subgroup selection decision and the selected test statistic which will be needed for calculation of Type 1 and Type 2 error rates.
At the interim analysis, let $W$ be the random variable which represents the decision about which subgroup has been selected (also called the population index). Let $w$ be the realisation of $W$ and this can take any value in the set $\Omega=\{1,2,F,\emptyset\}$. The notation $\emptyset$ indicates that it is possible to stop the trial for futility at the interim analysis without selecting a subgroup. Let $f_{Z^{(1)}_W|W}(z^{(1)}_w|W=w;\bs{\Theta})$ be the conditional distribution of the test statistic $Z_w^{(1)}$ given that $w$ has been selected. Then the joint probability density function is
$$
f_{Z^{(1)}_W, W}(z^{(1)}_w, w;\bs{\Theta})= \mathbb{P}(W=w;\bs{\Theta})f_{Z^{(1)}_W|W}(z^{(1)}_w|W=w;\bs{\Theta}).
$$ 
We note that the random variable $Z_\phi^{(1)}$ is not currently defined since if no subgroup is selected we cannot calculate a subgroup standardised statistic. However, it will be seen that the joint probability density function $f_{Z^{(1)}_W, W}(z^{(1)}_\phi, \phi;\bs{\Theta})$ is independent of $z_\phi^{(1)}$ and this joint probability function still has meaning. By Equation~\eqref{eq:CJD}, the test statistics are such that $Z^{(1)}_w\sim N(\theta_w\sqrt{I^{(1)}_w}, 1)$ for $w=1,2$ and $Z_1^{(1)}$ and $Z_2^{(1)}$ are independent. The conditional distribution $f_{Z^{(1)}_W|W}(z^{(1)}_w|W=w;\bs{\Theta})$ is given by a truncated normal distribution bounded below by $\zeta$. Hence, we have
\begin{align*}
&f_{Z^{(1)}_W, W}(z^{(1)}_1, 1;\bs{\Theta}) =\Phi\left(\zeta-\theta_2\sqrt{\mathcal{I}_2^{(1)}}\right)
\phi\left(z^{(1)}_1-\theta_1\sqrt{\mathcal{I}_1^{(1)}}\right) \\
&f_{Z^{(1)}_W, W}(z^{(1)}_2, 2;\bs{\Theta}) =\Phi\left(\zeta-\theta_1\sqrt{\mathcal{I}_1^{(1)}}\right)
\phi\left(z^{(1)}_2-\theta_2\sqrt{\mathcal{I}_2^{(1)}}\right) \\
&f_{Z^{(1)}_W, W}(z^{(1)}_F, F;\bs{\Theta}) =\frac{\sqrt{\mathcal{I}_1^{(1)}\mathcal{I}_2^{(1)}}}{\lambda(1-\lambda)\mathcal{I}_F^{(1)}} \\ &\times \int_{-\infty}^\infty \phi\left(\frac{\sqrt{\mathcal{I}_1^{(1)}}(u-\lambda\sqrt{\mathcal{I}_F^{(1)}})}{\lambda\sqrt{\mathcal{I}_F^{(1)}}}\right)
\phi\left(\frac{\sqrt{\mathcal{I}_2^{(1)}}(z_F^{(1)}-u-(1-\lambda)\sqrt{\mathcal{I}_F^{(1)}})}{(1-\lambda)\sqrt{\mathcal{I}_F^{(1)}}}\right) du \\
&f_{Z^{(1)}_W, W}(z^{(1)}_\phi, \phi;\bs{\Theta}) =\Phi\left(\zeta-\theta_1\sqrt{\mathcal{I}_1^{(1)}}\right)
\Phi\left(\zeta-\theta_2\sqrt{\mathcal{I}_2^{(1)}}\right)\\
\end{align*}
where $\phi(\cdot)$ and $\Phi(\cdot)$ denote the standard normal probability density and cumulative distribution functions respectively. We derive $f_{Z^{(1)}_W,W}(z^{(1)}_F, F;\bs{\Theta})$ in the supplementary materials.

\subsection{Calculation of Type 1 error and power}
\label{subsec:calc_errors}
We now consider the possible pathways of the enrichment trial. Then, given the definition of the $Z-$statistics, the threshold selection rule and the joint probability density function $f_{Z^{(1)}_W, W}(z^{(1)}_w, w;\bs{\Theta}), $ we are equipped to determine error rates for the study. We shall apply this method in Section~\ref{subsec:cond_score} in order to create an enrichment trial using the joint model for longitudinal and TTE data. The family wise error rate (FWER), denoted by $\alpha$, is defined as the probability of rejecting one or more true null hypotheses $H_j$ and power is denoted by $1-\beta$.

The testing procedure for this adaptive enrichment trial is described in Figure~\ref{fig:flow_chart}. At analysis $k$, let $(a_k,b_k)$ be an interval that splits the real line into three sections. We stop for futility if the test statistic of the selected subgroup, $Z^{(k)}_w,$ is below $a_k$, reject the corresponding null hypothesis and stop for efficacy if $Z^{(k)}_w$ is above $b_k$ and otherwise continue to analysis $k+1$. Let $H_G$ be the global null hypothesis, $\theta_1=\theta_2=\theta_F=0$. There are many pathways which lead to rejecting $H_G$. Examples include select $F$ and reject $H_{0,F}$ at the interim or select $S_1$ then reject $H_{0,1}$ at the final analysis. Considering all options, we have
\begin{equation}
\label{eq:type1}
\begin{split}
\alpha = \sum_{w\in \Omega}\bigg\{&\int_{b_1}^\infty f_{Z^{(1)}_W, W}\left(z^{(1)}_w, w;\bs{\Theta}_G\right)dz_w^{(1)} \\+&\int_{a_1}^{b_1}\int_{b_2}^\infty f_{Z_w^{(2)}|Z_w^{(1)}}\left(z_w^{(2)}|z_w^{(1)};\bs{\Theta}_G\right)dz_w^{(2)}dz_w^{(1)}\bigg\}.
\end{split}
\end{equation}
Here, we have specified that we will only test the hypothesis corresponding to the selected subgroup, since it has the highest chance of being significant. For alternative configurations testing all hypotheses, fixed sequence testing~\citep{westfall2001optimally} or other alpha propagation methods~\citep{tamhane2018gatekeeping} can be applied.

As in \citet{chiu2018design}, we define power as the conditional probability of rejecting $H_{0,1}$ given that subgroup $S_1$ is selected. Here, $S_1$ can be arbitrarily interchanged for $S_2$ or $F$. This reflects the belief that a ``successful" trial is one where the subgroup which benefits is selected and also reports a positive trial outcome. Following the same arguments as for Type 1 error, Type 2 error rates are calculated as
\begin{equation}
\label{eq:power}
\begin{split}
\beta = &\int_{-\infty}^{a_1} f_{Z^{(1)}_W, W}\left(z^{(1)}_1, 1;\bs{\Theta}_A\right)dz_1^{(1)} \\
+ &\int_{a_1}^{b_1}\int_{-\infty}^{a_2} f_{Z_1^{(2)}|Z_1^{(1)}}\left(z_1^{(2)}|z_1^{(1)};\bs{\Theta}_A\right)dz_1^{(2)}dz_1^{(1)}.
\end{split}
\end{equation}

It is now clear that the boundary points $a_1,a_2,b_1$ and $b_2$ can be calculated to satisfy pre-specified requirements of FWER $\alpha$, under $\bs{\Theta}_G,$ and power $1-\beta,$ under $\bs{\Theta}_A$. Further, to ensure that we have four equalities for the four boundary points, we make additional requirements that $\alpha^{(k)}$ is the Type 1 error ``spent" and $\beta^{(k)}$ is the Type 2 error spent at analysis $k$ where $\alpha^{(1)}+\alpha^{(2)}=\alpha$ and $\beta^{(1)}+\beta^{(2)}=\beta.$ Then solve
\begin{align*}
\alpha^{(1)} &= \sum_{w\in \Omega}\int_{b_1}^\infty f_{Z^{(1)}_W, W}\left(z^{(1)}_w, w;\bs{\Theta}_G\right)dz_w^{(1)} \\
\alpha^{(2)}&=\sum_{w\in \Omega}\int_{a_1}^{b_1}\int_{b_2}^\infty f_{Z_w^{(2)}|Z_w^{(1)}}\left(z_w^{(2)}|z_w^{(1)};\bs{\Theta}_G\right)dz_w^{(2)}dz_w^{(1)}\\
\beta^{(1)} &= \int_{-\infty}^{a_1} f_{Z^{(1)}_W, W}\left(z^{(1)}_1, 1;\bs{\Theta}_A\right)dz_1^{(1)}\\
\beta^{(2)}&= \int_{a_1}^{b_1}\int_{-\infty}^{a_2} f_{Z_w^{(2)}|Z_w^{(1)}}\left(z_1^{(2)}|z_1^{(1)};\bs{\Theta}_A\right)dz_1^{(2)}dz_1^{(1)}.
\end{align*}
The decomposition of the error rates also ensures that the boundary points $a_1$ and $b_1$ can be calculated at the first analysis before observing the information levels at the second analysis. Hence, there may be the opportunity to stop the trial early without needing to calculate the information levels at the second analysis. This is particularly helpful in trials which use TTE endpoints as information levels are estimated using the data.

There are many options for the break-down of the error rates. For the models considered, we shall use an error spending design~\citep{gordon1983discrete}. In the group sequential setting (without subgroup selection), the error spending test requires specifying the maximum information $\mathcal{I}_{max}$ and then error is spent according to the proportion of information $\mathcal{I}^{(k)}/\mathcal{I}_{max}$ observed at analysis $k$. For the enrichment trial, we propose a similar structure considering $\mathcal{I}_{max}$ to be the maximum information in the full population. Specifically, we shall use the functions $f(t) = \min\{\alpha t^2, \alpha\}$ and $g(t) = \min\{\beta t^2,\beta\}$ to determine the amount of error to spend. Then we set
\begin{align*}
\alpha^{(1)} &= f\left(\mathcal{I}_F^{(1)}/\mathcal{I}_{max}\right) \\
\alpha^{(2)} &= f\left(\mathcal{I}_F^{(2)}/\mathcal{I}_{max}\right)-f\left(\mathcal{I}_F^{(1)}/\mathcal{I}_{max}\right) \\
\beta^{(1)} &= g\left(\mathcal{I}_F^{(1)}/\mathcal{I}_{max}\right) \\
\beta^{(2)} &= g\left(\mathcal{I}_F^{(2)}/\mathcal{I}_{max}\right)-g\left(\mathcal{I}_F^{(1)}/\mathcal{I}_{max}\right)
\end{align*}
We shall discuss the calculation of $\mathcal{I}_{max}$ in the TTE (or joint modeling) setting in Section~\ref{subsec:design}.

By construction, under $H_G:\theta_1=\theta_2=\theta_F=0$, we have FWER $\alpha$ exactly by Equations~\eqref{eq:type1} and~\eqref{eq:power}. Hence, the FWER is protected in the weak sense. To prove that we also have strong control of the FWER, we impose the condition that the treatment effect in the full population, is non-negative. This ensures that the subgroup selected does not differ under scenarios $(\theta_1,\theta_2)$ and $(0,0)$ which is needed for the proof. The condition is not restrictive, since treatment effects other than $\theta_F$ are allowed to be negative and $\theta_F$ can equal zero.
\begin{theorem}
\label{theorem:strong_control}
For global null hypothesis $H_G$ and any $\bs{\Theta}=(\theta_1,\theta_2)$ such that $\theta_F=\lambda \theta_1+(1-\lambda)\theta_2$ is non-negative, we have
\[ \mathbb{P}(\text{Reject at least one true }H_j|\bs{\Theta})\leq \mathbb{P}(\text{reject at least one }H_j|H_G).\]
\end{theorem}
\begin{proof}See the supplementary materials.
\end{proof}
In Section~\ref{sec:results}, we also show by simulation, that the FWER is protected at significance level $\alpha=0.025$ and is not conservative.
 
\subsection{Trials with unpredictable information increments: events based analyses}
\label{subsec:information_predict}
To complete the calculation of the boundary points $a_2$ and $b_2$ in Equations~\eqref{eq:type1} and~\eqref{eq:power}, it remains to find the information level at analysis $2$ for the subgroups that have ceased to be observed. That is, suppose that $w\in\{1,2,F\}$ is the subgroup that has been selected and the trial continues to analysis $2,$ then $\mathcal{I}_w^{(2)}$ is observed. However, we also need to know $\mathcal{I}_j^{(2)}$ for all $j=1,2,F$ such that $w\neq j$. Many enrichment trial designs focus on the simple example where the outcome measure is normally distributed with known variance. Hence, if the number of patients to be recruited is pre-specified, then information levels can be calculated in advance of the trial and this problem does not occur. However, in trials where the primary endpoint is a TTE variable, information is estimated using the data. We find that we can accurately forward predict the information levels at future analyses when we know the number of observed events. Hence, to mitigate the problem of not knowing $\mathcal{I}_j^{(2)}$, we shall pre-specify the number of observed events.

For subgroup $j=1,2$, let $d_j^{(k)}$ be the number of events observed in subgroup $j$ by analysis $k$. We plan that if no early stopping occurs, then the total number of observed events in the selected subgroup is the same regardless of which subgroup has been selected so that $d^{(2)}_1=d^{(2)}_2=d^{(2)}_F=d^{(2)}$. Figure~\ref{fig:flow_chart} identifies when the analyses are performed.  Note that these values are set as design options and so will be known before commencement of the trial. We shall discuss how to choose these values in Section~\ref{subsec:design}. 

\begin{figure}[t]
\centering
\includegraphics[width=0.7\textwidth]{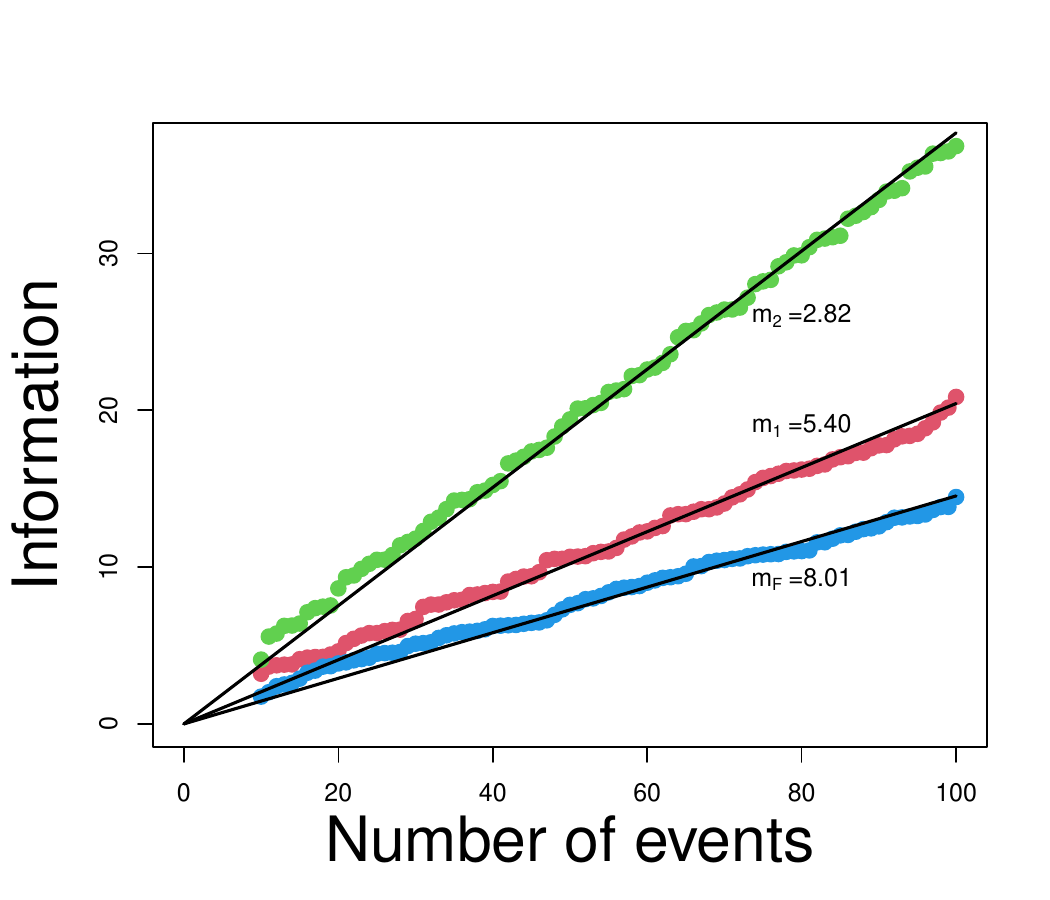}
\caption{Calculation of constants \(m_1,m_2\) and \(m_F\). Result shows that information is proportional to number of events.\label{fig:inf_proportional}}
\end{figure}

Further, we relate number of events and information so that we can predict the information level at the second analysis for the unobserved subgroups. \citet{freedman1982tables} proves that, in the context of survival analysis, the variance of the log-rank statistic under $H_G$ is such that $\mathcal{I}^{(k)}_j \approx d^{(k)}_j/4$. For analysis methods using test statistics other than the log-rank, we shall extend on this idea and assume that $\mathcal{I}^{(k)}_j = d^{(k)}_j/m_j$ where $m_j$ is a constant. Figure~\ref{fig:inf_proportional} shows evidence that the assumed relationship between number of events and information holds.

For now, we need only the assumption of the structural form of this relationship.  At the interim analysis, each $\mathcal{I}^{(1)}_j$ is observed for $j=1,2,F.$ Hence, we can use the proportionality relationship to predict the information at the second analysis for the subgroup which is no longer observed. For $j\neq w$ we can predict $\mathcal{I}_j^{(2)}$ using
$$
\mathcal{I}_j^{(2)}= d_j^{(2)} \frac{1}{m_j} = d^{(2)}\frac{\mathcal{I}_j^{(1)}}{d^{(1)}_j} \hspace{1cm} \text{for } j=1,2,F.
$$

\subsection{Trial design --- number of events}
\label{subsec:design}
We have so far presented the calculation of the boundary points for a trial where the number of events at the interim and final analyses are known prior to commencement. We now discuss the design of the trial, in particular, determining the constants $m_j$ and information levels $\mathcal{I}_j^{(1)}$ for $j=1,2,F$ and maximum information level $\mathcal{I}_{max}$. These in turn mean that the required numbers of events $d^{(1)}_j$ for $j=1,2,F$ and $d^{(2)}$ can be planned. The driving design feature is that we will plan the trial to have power $1-\beta$ under the parameterisation $\bs{\Theta}_A$. We now describe a simulation scheme to determine the constants $m_j$ for $j=1,2,F.$
\begin{algorithm}
\begin{algorithmic}\
\State Under the parameterisation $\bs{\Theta}_A$, simulate a data set of $5000$ patients
\State  Let $t_{j,1},\dots, t_{j,n_j}$ be the event times in subgroup $j$
\For{$t_{j,s} = t_{j,1},\dots, t_{j,n_j}$} 
    \State  Right-censor all patients at time $t_{j,s}$
    \State Calculate $\mathcal{I}_{j,s}^{(1)}$ based on data up to time $t_{j,s}$
\EndFor
\State Fit a linear model, without an intercept term, to the points $(t_{j,1},\mathcal{I}_{j,1}^{(1)}),\dots,(t_{j,n_j},\mathcal{I}_{j,n_j}^{(1)})$ 
\State Use this linear model to estimate the value of $m_j$.
\end{algorithmic}
\end{algorithm}

Figure~\ref{fig:inf_proportional} gives a graphical representation of this scheme. It is now possible to calculate the required number of events at the first interim analysis. In the example in Section~\ref{subsec:threshold}, we require $\mathcal{I}_1^{(1)}=9.08$ which equates to $d_1^{(1)}=9.08m_1$ events in subgroup $S_1$. Further, we find that $m_2=(1-\lambda)m_1/\lambda$ and $m_F=m_1/\lambda$ which equates to $d_2^{(1)}=(1-\lambda)d_1^{(1)}/\lambda$ and $d_F^{(1)}=d_1^{(1)}/\lambda$ and this can be seen in Figure~\ref{fig:inf_proportional}. The design of the trial does not require us to plan $d^{(1)}_2$ and $d^{(1)}_F$, but this provides us with estimates of the number of events that will be observed at the first analysis. We can also determine the timing of the final analysis at $K=2$. Consider the sequence of information levels given by
$$
(\tilde{\mathcal{I}}^{(1)}_j, \tilde{\mathcal{I}}^{(2)}_j) = \left(d_j^{(1)}/m_j, m_F\mathcal{I}_{max}/m_j\right)
$$
for $j\in =1,2,F.$ The value of $\mathcal{I}_{max}$ is calculated such that boundary points satisfy $a_K=b_K$ when the information levels $\tilde{\mathcal{I}}_j^{(k)}$ replace $\mathcal{I}_j^{(k)}$ in Equations~\eqref{eq:type1} and~\eqref{eq:power} for $k=1,2$ and $j=1,2,F.$ This is done using an iterative search method. Then, returning to the definition of $\mathcal{I}_{max}$, the total number of events can be found by solving $\mathcal{I}_F^{(2)} = \mathcal{I}_{max}$ for $d^{(2)}$. In Section~\ref{sec:results}, we present the sample sizes which have been calculated for a range of parameter choices.
\section{Alternative models and their analysis methods}
\label{sec:survival}
\subsection{Cox proportional hazards model}
\label{subsec:cox}
Methods which leverage information from biomarkers in TTE data in enrichment trials are yet to be established. The current best practice for adaptive designs with a TTE endpoint is to base analyses on Cox proportional hazards models. We emulate this conventionality in order to assess the gain from including the longitudinal data in the analysis. To do so, we shall present a simple Cox proportional hazards model and define treatment effect estimates that can be used in accordance with the threshold selection rule to perform an enrichment trial.

Denote $h_{0j}(t)$ as the baseline hazard function, $\theta_j$ the treatment parameter and $\psi_{ji}$ as the treatment indicator that patient $i$ in subgroup $j=1,2$  receives the new treatment. Then the hazard function for the survival model is given by
\begin{equation}
\label{eq:cox_model}
h_{ji}(t)=h_{0j}(t)\exp\{\theta_j \psi_{ji}\}.
\end{equation}
We note the similarities and differences between this model and the joint model of Section~\ref{subsec:joint_model}. In the results that follow in Section~\ref{subsec:sim_study}, we shall assume that the joint model is true (and simulate data from the joint model). However, we fit the data to the Cox proportional hazards model which highlights that this will be a misspecified model.

When analysing data using this model, the null hypothesis in Equation~\eqref{eq:hypothesis} can be tested at analysis $k=1,\dots,K$ by calculating treatment effect estimates $\hat{\theta}_j^{(k)}$, information levels $\mathcal{I}_j^{(k)}$ and $Z$-statistics for $j=1,2,F$. As described in the supplementary materials, $\hat{\theta}_j^{(k)}$ is given as the root of the equation where the partial score statistic is set equal to zero~\citep{jennison1997group} and the information $\mathcal{I}_j^{(k)}$ as the first derivative of the partial score statistic. \citet{jennison1997group} prove that the resulting $Z$-statistics have the CJD given in Equation~\eqref{eq:CJD} and the methodology of Sections~\ref{sec:enrichment_theory} can be used to create an enrichment trial design. 

\subsection{Cox proportional hazards model with longitudinal data as a time-varying covariate}
\label{subsec:cox_time_varying}
A final option for analysis is one where the longitudinal data is included but is assumed to be free of measurement error. This requires a more sophisticated model than the simple Cox proportional hazards model of Section~\ref{subsec:cox} and represents a trial where the longitudinal data is regarded as important enough to be considered and included in the model. However, this is still a naive approach since the model will be misspecified in the presence of measurement error. For the purpose of assessing the necessity of correctly modeling the data, we shall fit a Cox proportional hazards model to the data where the longitudinal data is treated as a time-varying covariate.

In what follows, the definitions of the treatment indicator $\psi_{ji}$ and longitudinal data measurements $W_{ji}(v_{ji1}),\dots,W_{ji}(v_{jim_{ji}})$ remain the same as in Section~\ref{subsec:joint_model}. Let $\gamma_j$ and $\theta_j$ be longitudinal data and treatment parameters respectively, then the hazard function is given by
\begin{equation}
\label{eq:cox_long}
h_{ji}(t)=h_{j0}(t)\exp\{\gamma_jW_{ji}(t) + \theta_j\psi_{ji} \}.
\end{equation}
This model differs from the joint model because the assumption here is that $W_{ji}(t)$ is a function of time that is measured without error. In reality we often have measurements $W_{ji}(v_{ji1}),\dots,W_i(v_{jim_{ji}})$ for patient $i$ in subgroup $j$ that include noise around a true underlying trajectory.

In a similar manner to Section~\ref{subsec:cox}, the hypothesis in Equation~\eqref{eq:hypothesis} can be tested by finding $Z-$statistics, with the CJD of Equation~\eqref{eq:CJD}~\citep{jennison1997group} and following the enrichment trial design of Section~\ref{sec:enrichment_theory}.
\section{Results}
\label{sec:results}
\subsection{Simulation set-up}
In what follows, we perform simulation studies to assess the Type 1 error rates and observed power for the three analysis methods of Sections~\ref{sec:joint} and~\ref{sec:survival}. These methods shall herto be referred to as ``Conditional score", ``Cox" and ``Cox with biomarker" respectively. The purpose of this comparison is to assess the gain by including the longitudinal data and to decide whether correctly modeling the measurement error is necessary.

For the presented analyses, we shall assume that the joint model is true. Hence, the working model for data generation is given by Equations~\eqref{eq:long1}--\eqref{eq:surv}. Each of the analysis methods have the advantage that we need not specify the baseline hazard function since each method is semi-parametric and requires no assumptions regarding $h_{0j}(t).$ Even when the method includes the longitudinal data, there are no distributional assumptions about the random effects $\mathbf{b}_{j1},\dots,\mathbf{b}_{jn_j},$ ensuring it is robust to some model misspecifications. For the purpose of simulation however, we now describe the distributions used for data generation. We shall simulate data with baseline hazard function given by
\begin{equation}
\label{eq:baseline_hazard}
h_{0j}(t) =\begin{cases}
   c_{j1} &\text{ if } t \leq 1 \\
   c_{j2} &\text{ if } t > 1
\end{cases}.
\end{equation}
We have chosen to simulate from a model where the baseline hazard function as piece-wise constant with a single knot-point at time $t=1$ for simplicity. This is motivated by the metastatic breast cancer data where we see a sharp difference in the baseline hazard at 1 year. It is straight forward to extend this to a general piece-wise constant baseline hazard function with multiple knot-points. We consider a random effects model where $\mathbf{b}_{j1},\dots,\mathbf{b}_{jn}$ are independent and identically distributed with the following distribution
\begin{equation}
\label{eq:random_effects}
\begin{bmatrix}b_{0ji} \\b_{1ji} \end{bmatrix} \sim N\left(\begin{bmatrix} \mu_{1j} \\ \mu_{2j} \end{bmatrix} , \begin{bmatrix} \phi_{1j} & \phi_{12j} \\ \phi_{12j} & \phi_{2j} \end{bmatrix} \right).
\end{equation}

The parameter values for simulation studies are informed using the metastatic breast cancer dataset~\citep{dawson2013analysis}. We removed patients whose ER status is negative and measurements of ctDNA which were ``not detected" were set to 1.5 (Copies/ml)~\citep{barnett2021methods}. The dataset contains multiple treatment arms and dosing schedules, hence, we usethis dataset to represent standard of care (control group). The parameter values, which have been suitably rounded, shall remain fixed throughout the simulation studies are given by
\begin{equation}
\begin{split}
&\lambda=2/3, \gamma = \gamma_1=\gamma_2=0.8, \\
&(\phi_{1},\phi_{12},\phi_{2}) = (\phi_{11},\phi_{121},\phi_{21})=(\phi_{12},\phi_{122},\phi_{22})=(2.5,1.7,5), \\
&\sigma^2 = \sigma_1^2=\sigma_2^2=0.25,(\mu_{01},\mu_{11})=(\mu_{02},\mu_{12})=(4.23, 1.81), \\
&c_{11}=c_{21}=0.0085,c_{12}=c_{22}=0.0142.
\end{split}
\label{eq:values}
\end{equation}
We shall perform simulation studies for a range of $\gamma,\sigma^2$ and $\phi_2$ values. The interpretation of these parameter are now described. $\gamma$ describes the association between the biomarker and time-to-event outcomes. Higher values of $\gamma$ lead to higher correlation between the two endpoints. The parameter $\sigma^2$ controls the noise in the measurement error of the longitudinal data. Finally, $\phi_2$ represents the variance of the slopes of the random effects terms and therefore the degree of similarity between patients' longitudinal trajectories. 

For our simulations, patients are recruited at a rate of 2 per week so that enrollment is slow and adaptive methods are appropriate. The recruitment ratio of control to experimental treatment is fixed as 1:1 for all subgroups and all simulations studies. ctDNA observations will be collected, via a blood test, at two weeks for the first three months following entry to study and then once per month. The final object of importance which is required for data generation is the mechanism which simulates censoring times, $y_1,\dots,y_n$. We shall simulate these according to an exponential distribution with rate parameter $5\times 10^{-5}$ (years) and this is independent of the time-to-event outcome to reflect non-informative censoring. This results in roughly $10\%$ of patients being lost to follow-up.

To complete the set-up, we now present the sample sizes used for each simulation study and these values have been calculated by employing the methods of Section~\ref{subsec:design}. The trial is planned with FWER $\alpha=0.025$ and planned power $1-\beta=0.9$. The number of events at the first analysis in subgroup $S_1$, denoted $d^{(1)}_1$, have been chosen to ensure that subgroup $S_1$ is selected roughly $60\%$ of the time and the total number of events at the second analysis, $d^{(2)},$ have been chosen to attain power of $90\%$ as described in Section~\ref{subsec:design}. In all cases, the value of $d^{(1)}_1$ is large enough such that the survival data is mature at the interim analysis and decisions can be made with confidence. These numbers of events are displayed in Table~\ref{tbl:design} for a range of values of $\gamma,\sigma^2$ and $\phi_2$. As $\gamma$ increases, we see that required $d^{(1)}_1$ and $d^{(2)}$ increase. Similarly, the required number of events increase with $\sigma^2.$ That is, more events and hence more information is needed to achieve power and selection probabilities when the longitudinal data is noisy. When $\sigma^2=2.25$ and with a small number of events at the first interim analysis, it is not always possible to find a root to Equation~\eqref{eq:score_1}. Consequently, the required $d^{(1)}_1$ and $d^{(2)}$ are high to ensure that large sample properties of the estimator hold. We have not seen this problem occur for $\sigma^2\leq 2.25.$ The values of $d_1^{(1)}$ and $d^{(2)}$ appear to be immune to changes in $\phi_2$.
\begin{table}
\caption{Sample size calculations for the adaptive enrichment trial. $d^{(1)}_1$
is the required number of events in subgroup $S_1$ at the interim analysis
and $d^{(2)}$ is the total number of events in the selected subgroup at the final analysis.
Number of events calculated to satisfy FWER 0.025 and power 0.9.}
\label{tbl:design}
\begin{center}
\begin{tabular}{|lll|cc|}
\hline
$\bs{\gamma}$ & $\bs{\sigma}^2$ & $\bs{\phi}_2$ & $d^{(1)}_1$ & $d^{(2)}$
\\\hline
\rule{0pt}{3.5ex}$\mathbf{0}$ & 
     $\mathbf{0.25}$ &
     $\mathbf{5}$ &40 &174\\
$\mathbf{0.4}$ & 
     $\mathbf{0.25}$ &
     $\mathbf{5}$ &47 &204\\
$\mathbf{0.8}$ & 
     $\mathbf{0.25}$ &
     $\mathbf{5}$ &47 &206\\
$\mathbf{1.2}$ & 
     $\mathbf{0.25}$ &
     $\mathbf{5}$ &50 &218\\
\rule{0pt}{3.5ex}$\mathbf{0.8}$ & 
     $\mathbf{0}$ &
     $\mathbf{5}$ &45 &194\\
$\mathbf{0.8}$ & 
     $\mathbf{0.25}$ &
     $\mathbf{5}$ &47 &206\\
$\mathbf{0.8}$ & 
     $\mathbf{1}$ &
     $\mathbf{5}$ &58 &252\\
$\mathbf{0.8}$ & 
     $\mathbf{2.25}$ &
     $\mathbf{5}$ &69 &301\\
\rule{0pt}{3.5ex}$\mathbf{0.8}$ & 
     $\mathbf{0.25}$ &
     $\mathbf{0}$ &46 &198\\
$\mathbf{0.8}$ & 
     $\mathbf{0.25}$ &
     $\mathbf{2.5}$ &44 &194\\
$\mathbf{0.8}$ & 
     $\mathbf{0.25}$ &
     $\mathbf{5}$ &47 &206\\
$\mathbf{0.8}$ & 
     $\mathbf{0.25}$ &
     $\mathbf{7.5}$ &47 &203\\
\hline
\end{tabular}
\end{center}
\end{table}

\subsection{Type 1 error rate comparison}
The first important comparison will be the Type 1 error rate using each of the analysis methods Conditional score, Cox and Cox with biomarker.

\begin{figure}[htb]
\begin{subfigure}[t]{0.49\textwidth}
\includegraphics[width=\textwidth]{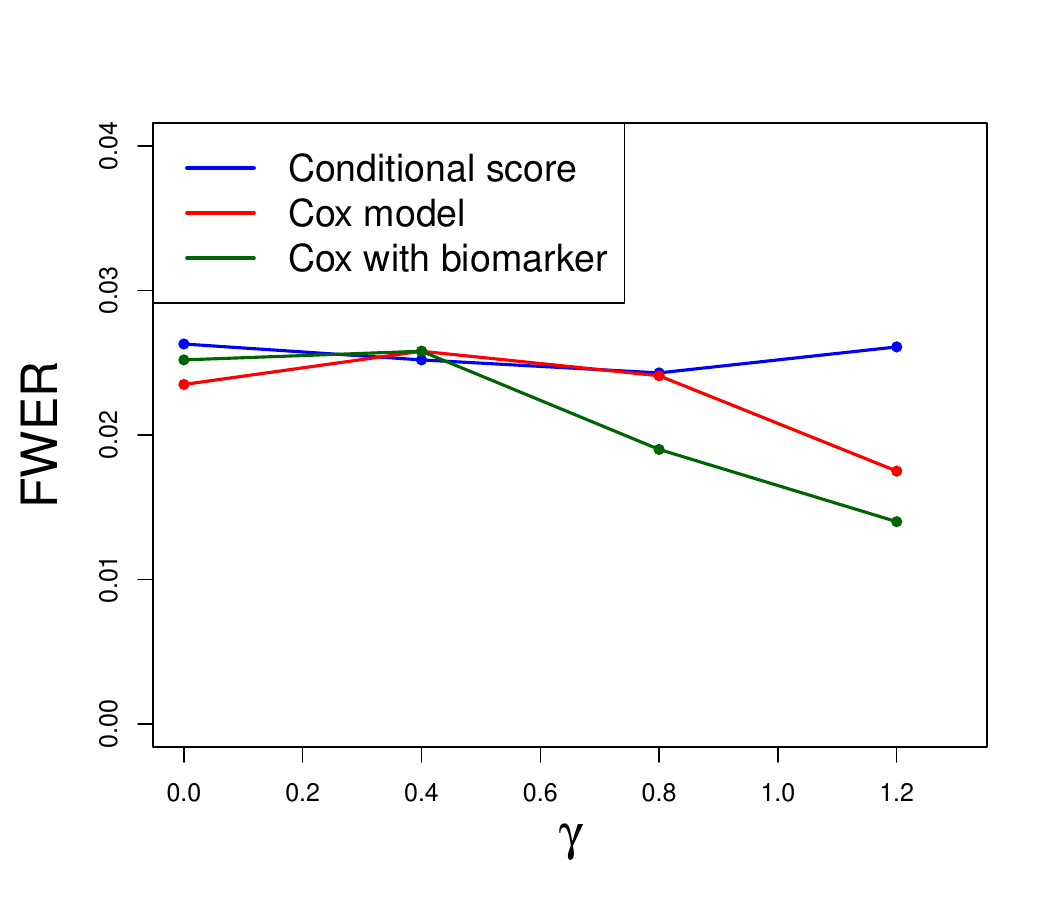}
\end{subfigure}
\hfill
\begin{subfigure}[t]{0.49\textwidth}
\includegraphics[width=\textwidth]{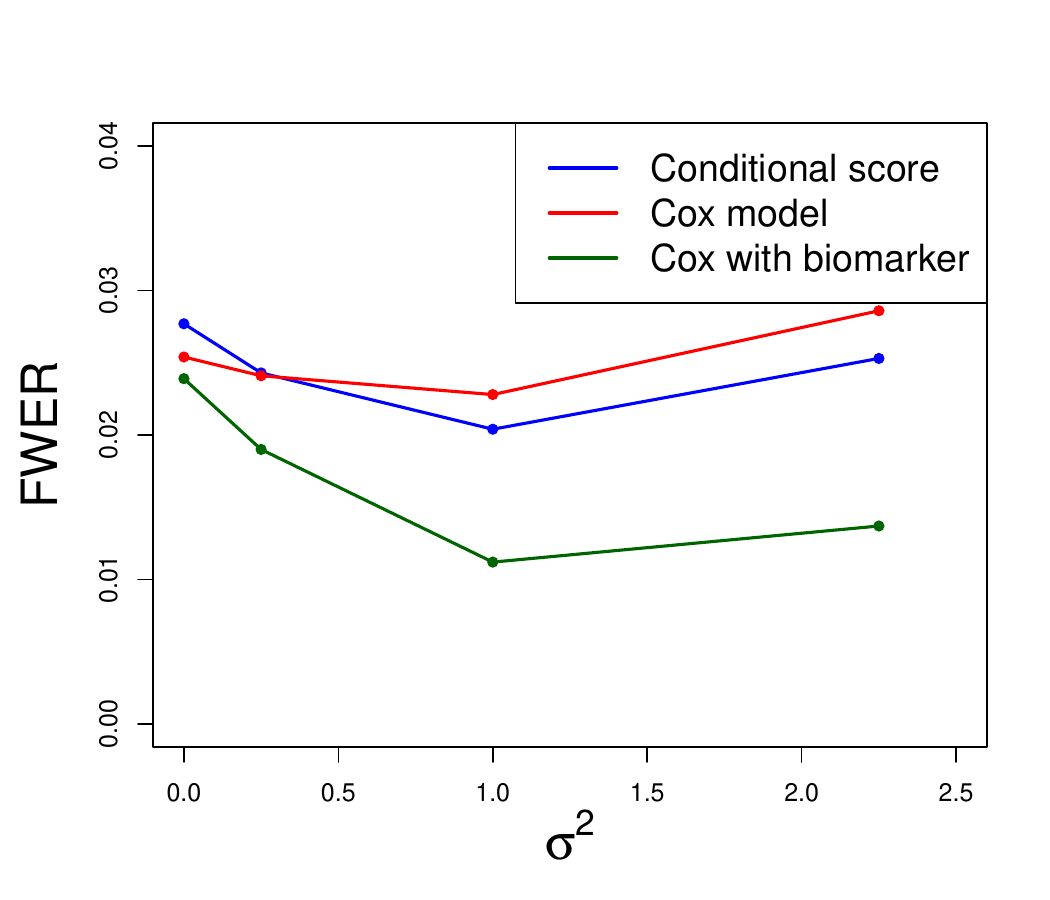}
\end{subfigure}
\centering
\begin{subfigure}[t]{0.49\textwidth}
\includegraphics[width=\textwidth]{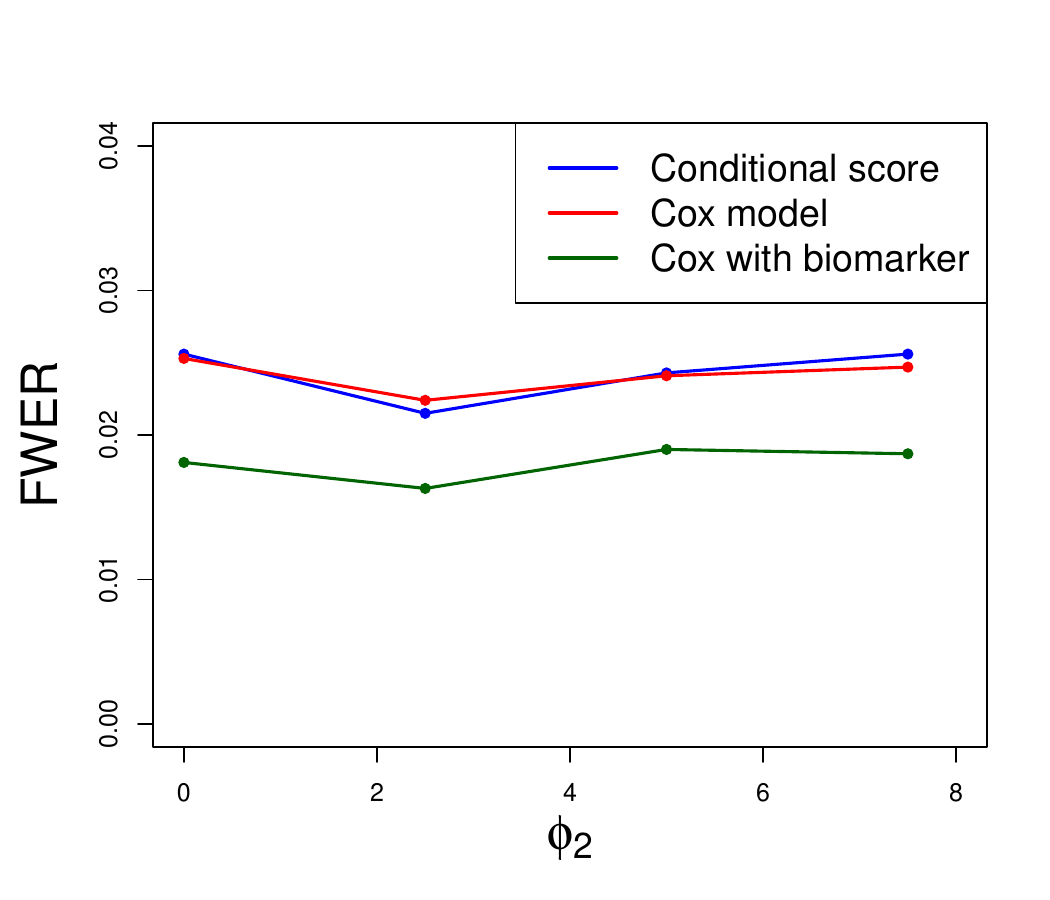}
\end{subfigure}
\caption{Type 1 error rates displaying changes in parameters $\gamma,\sigma$ and $\phi_2$. All other parameters are as in~\eqref{eq:values}. Numeric values of the points are presented in the supplementary materials. For a study with $N=10^4$ simulations and FWER 0.025, simulation standard error is 0.00156.}
\label{fig:type1}
\end{figure}

To represent no differences between control and treated groups under $H_{0j}$, let $\theta_j=0$ for each $j=1,2,F$. Figure~\ref{fig:type1} shows the results of a simulation study assessing the FWER for each method and different parameter values. For each simulation, a dataset of patients is generated from the joint model, then subgroup selection and decisions about $H_0$ are performed after $d_1^{(1)}$ and $d^{(2)}$ events have been observed according to Table~\ref{tbl:design}. All four methods are performed on the same dataset and after the same number of events so that differences can be attributed to the analysis methodology and not trial design features.

It is clear that for the majority of cases, the FWER is controlled when the Conditional score method is used to estimate the treatment effect in the joint model. For a study with $N=10^4$ simulations and planned significance value $\alpha=0.025$, the simulation error bounds is $(0.0219,0.0281)$. Hence, the observed FWER is within reasonable distance from $\alpha=0.025$ in accordance with the number of simulations. The result of Theorem~\ref{theorem:strong_control} together with the simulation result in Figure~\ref{fig:type1} give us confidence that FWER is controlled at the desired significance level using the joint modeling approach. The Cox model also appears to control the FWER but may be seen to be conservative for large values of $\gamma.$. However, we see that the Cox with biomarker method has FWER considerably smaller than 0.025. This is particularly apparent for $\sigma^2 \geq 1$ and all values of $\phi_2.$

\subsection{Efficiency comparison}
\label{subsec:sim_study}
\begin{figure}[htb]
\begin{subfigure}[t]{0.49\textwidth}
\includegraphics[width=\textwidth]{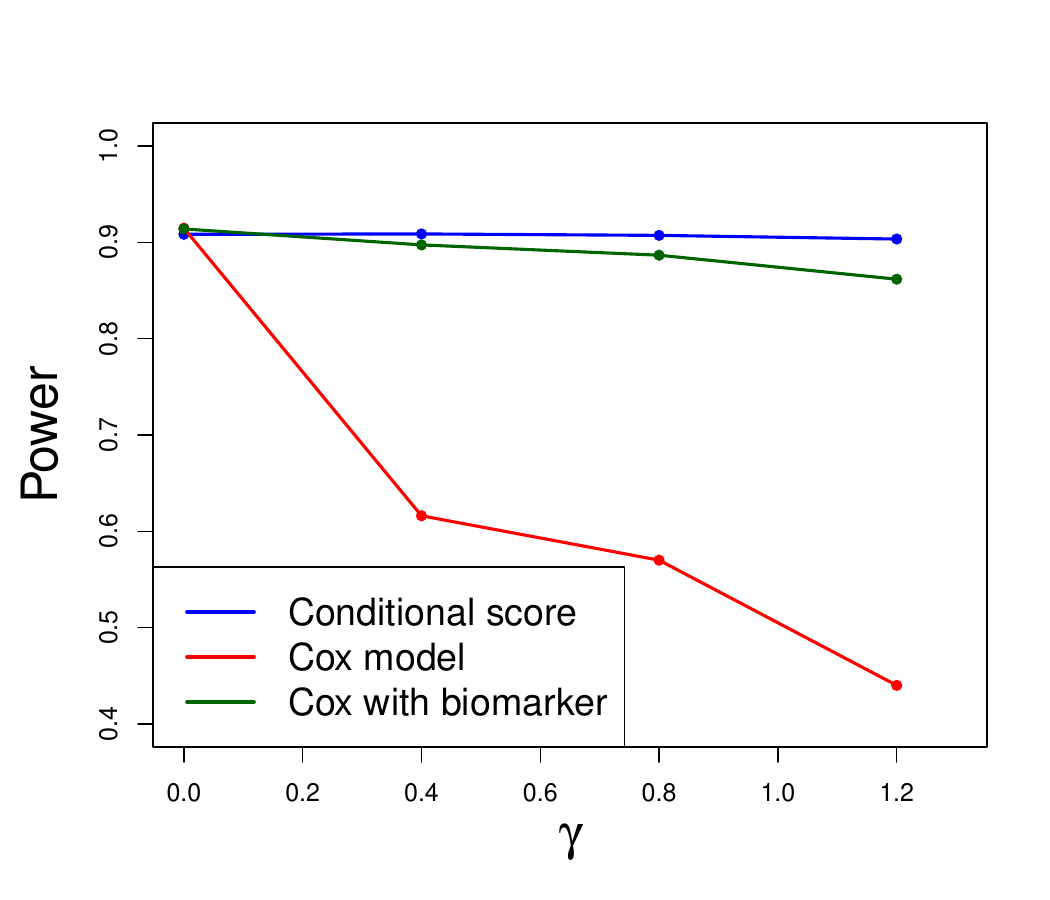}
\end{subfigure}
\hfill
\begin{subfigure}[t]{0.49\textwidth}
\includegraphics[width=\textwidth]{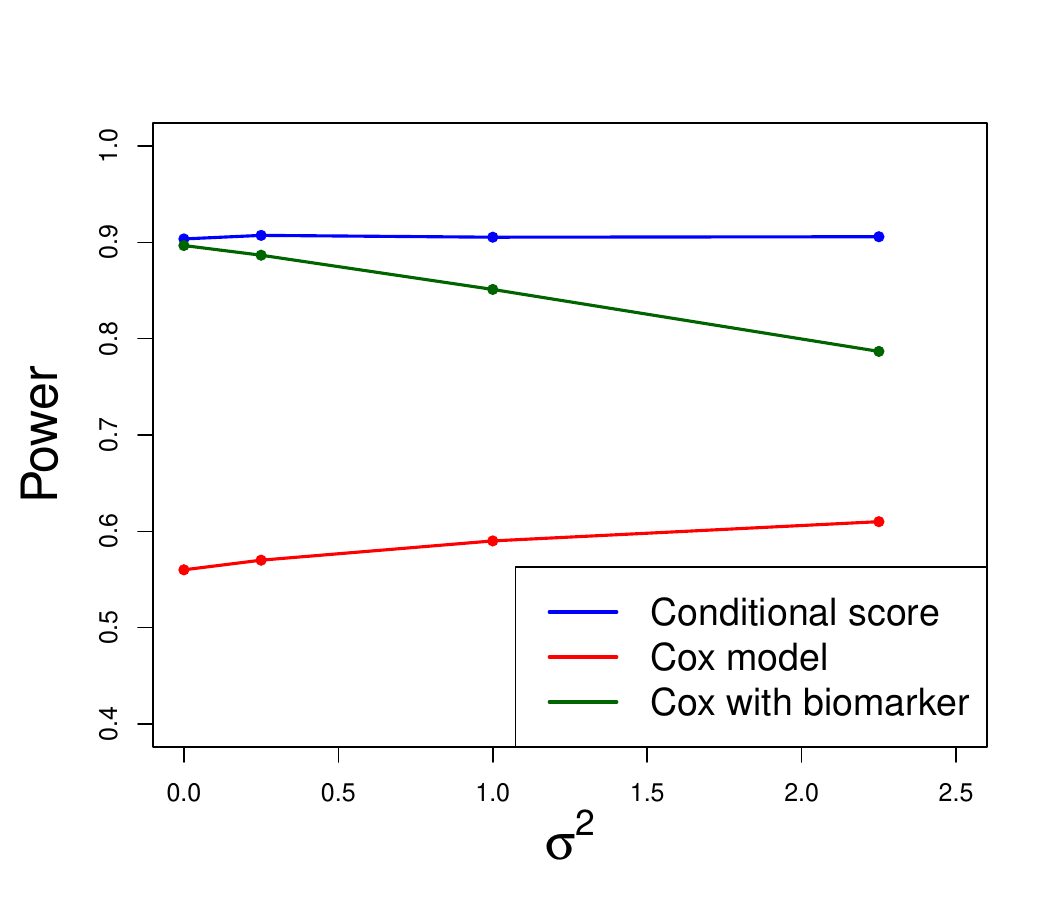}
\end{subfigure}
\centering
\begin{subfigure}[t]{0.49\textwidth}
\includegraphics[width=\textwidth]{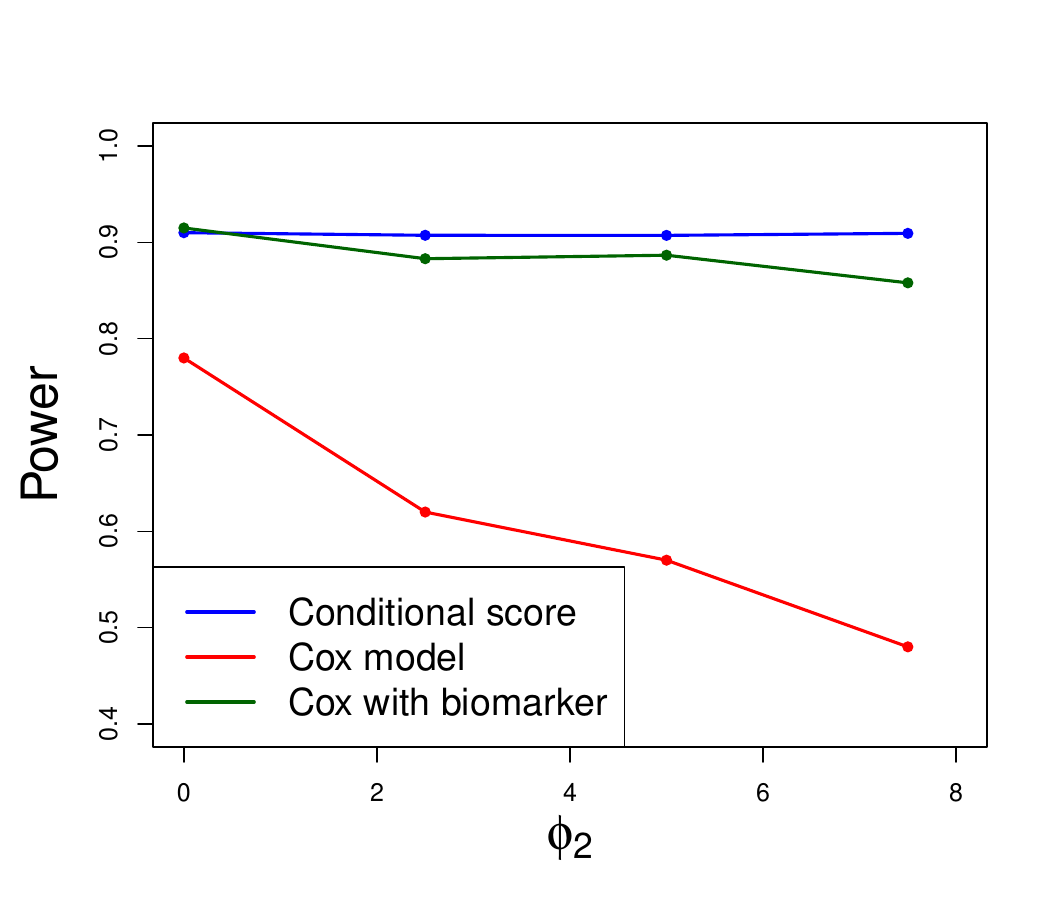}
\end{subfigure}
\caption{Observed power displaying changes in parameters $\gamma,\sigma$ and $\phi_2$. All other parameters are as in~\eqref{eq:values}. Numeric values of the points are presented in the supplementary materials. For a study with $N=10^4$ simulations and power 0.9, simulation standard error is 0.003.}
\label{fig:power}
\end{figure}

We shall focus on power as a measure of efficiency between the different methods and we compare some other outcome measures, such as number of hospital visits and expected stopping time in the supplementary materials. Under the alternative, only patients in subgroup $S_1$ will respond to treatment, represented by $H_{A1}:\theta_1=-0.5$ and $H_{A2}:\theta_2=0.$ Figure~\ref{fig:power} shows the power comparison between the different methods. Power is calculated as the proportion of simulations which reject $H_{01}$ out of those where subgroup $S_1$ is selected, as described in Section~\ref{subsec:calc_errors}. 

It is clear that the conditional score method is most efficient since power is highest across nearly all parameter combinations. When $\gamma=0$, the conditional score method may suffer from a small loss in power in comparison to other methods. This is the case where longitudinal data has no impact on the survival outcome so including it in the analysis is futile. For $\gamma \neq 0$ however, a gain in power up to 0.46 is seen.

Fitting the data to the simple Cox model is very inefficient and in the extreme cases, power is below $0.5.$ The sample size that would be needed to increase power to 0.9 in such a scenario is excessive. This simple method has power lower than the conditional score method whenever $\gamma\neq 0$ and becomes increasingly inefficient as $\gamma$ increases and as $\phi_2$ increases. The efficiency of this method appears to increase slightly with $\sigma.$ Hence, it is important to include the longitudinal data in the analysis when there is a suspected correlation between the longitudinal data and the survival endpoint.

The final method, where TTE outcomes are fit to a Cox proportional hazards model with the longitudinal data as a time-varying covariate, appears to be a simple yet effective way of including longitudinal data in the analysis. The achieved power is at least 0.78 but is usually lower than the conditional score method. However, scenarios where this method outperformes the conditional score are when $\sigma=0$ or $\phi_2=0$ indicating that the longitudinal data is free of measurement error or there are no between-patient differences in the slopes of the longitudinal trajectories. The efficiency decreases as longitudinal data increase in noise or as patient differences become larger, that is as $\sigma$ and $\phi_2$ increase.

An advantage of the two alternative Cox models is that there is no criteria to have a minimum of two longitudinal observations to be included in the at-risk process. In fact, for these alternative models, we need not specify the functional form of the trajectory of the longitudinal data, for example that it is linear in time. Taking these considerations into account, we believe that the most efficient and practical method is the conditional score, which includes the longitudinal data and takes into account the measurement error.

\section{Discussion}
\label{sec:discussion}
We have shown that the threshold selection rule can be combined with an error spending boundary to create an efficient enrichment trial. This is potentially suitable for any trial where the primary outcome is a TTE variable and we present a method to establish the required number of events at the design stage of the trial. A novel aspect of this work is that these methods can be applied to an endpoint which is the treatment effect in a joint model for longitudinal and TTE data. By including these routinely collected biomarker outcomes in the analysis to leverage this additional information, the enrichment trial has higher power compared to the enrichment trial where the longitudinal data is left out of the analysis. \citet{bauer2010selection} show that bias is prevalent in designs with selection. In our case, selection bias occurs as the treatment effect estimate in the selected subgroup is inflated in later analyses which could affect the trial results. However, unlike most other selection schemes, the threshold selection rule adjusts for the magnitude of the treatment effect at the design stage so another advantage is that selection bias is incorporated into the decision making process. 

Further, we compared this joint modeling approach with a model which used the longitudinal data but naively assumed this was free of measurement error. Again, the joint model performed more effectively in most cases. This naive approach was more efficient when the longitudinal data was truly free from measurement error, there was no correlation between the two endpoints or there was no heterogeneity between patients' biomarker trajectories. However, we believe that these situations are rare in practice and the gain in power from joint modeling outweighs this downside.

\section*{Data availability statement}
All data is simulated according to the specifications described.

\section*{Funding}
This project has received funding from the European Union’s Horizon 2020 research and innovation programme under grant agreement No 965397. TJ also received funding from the UK Medical Research Council (MC\_UU\_00002/14). RB also acknowledges funding from Cancer Research UK and support for his early phase clinical trial work from the Cambridge NIHR Biomedical Research Centre (BRC-1215-20014) and Experimental Cancer Medicine Centre. For the purpose of open access, the author has applied a Creative Commons Attribution (CC BY) licence to any Author Accepted Manuscript version arising.

\section*{Declaration of conflicting interests}
The authors declare no potential conflict of interests.

\section*{ORCID IDs}
Abigail Burdon https://orcid.org/0000-0002-0883-4160 \\
Richard Baird https://orcid.org/0000-0001-7071-6483 \\
Thomas Jaki https://orcid.org/0000-0002-1096-188X

\section*{Supplementary material}
Supplementary material for this article is available online.

\noindent Software in the form of R code, is available at https://github.com/abigailburdon/Adaptiveenrichment-with-joint-models.

\bibliographystyle{plainnat}  
\bibliography{enrichment}

\end{document}


\maketitle

\section{Alternative joint model and estimation method}
\subsection{Joint model with treatment effect on the biomarker}
As described in the main text, it may be desirable to have a joint model which accommodates a treatment effect in the biomarker trajectory. This model and analysis method are now described.

Let the times of the measurements of the longitudinal data for patient $i$ in subgroup $j=1,2$ be denoted by $v_{ji1},\dots v_{jim_{ji}}$, then $X_{ji}(v_{jis})$ is the true value of the biomarker at time $v_{jis}$ and $W_{ji}(v_{jis})$ is the observed value of the biomarker. Suppose that $\psi_{ji}$ is the indicator function that patient $i$ in subgroup $j=1,2$ receives the experimental treatment. Let $\theta_j$ and $\gamma_j$ be scalar coefficients. The longitudinal data model takes the form
\begin{equation}
\begin{split}
\label{eq:joint}
X_{ji}(v_{jis})&=b_{0ji}+b_{1ji}v_{jis}+b_2v_{jis}\psi_{ji} \\
W_{ji}(v_{jis})&= X_{ji}(v_{jis})+\epsilon_{ji}(v_{jis}) \\
\begin{bmatrix}b_{0ji} \\b_{1ji} \end{bmatrix} &\sim N\left(\begin{bmatrix} \mu_{1j} \\ \mu_{2j} \end{bmatrix} , \begin{bmatrix} \phi_{1j} & \phi_{12j} \\ \phi_{12j} & \phi_{2j} \end{bmatrix}\right)
\end{split}
\end{equation}
where $\mathbf{b}_{ji}=(b_{0ji},b_{1ji})$ is a vector of patient specific random effects and $\epsilon_{ji}(v_{jis})$ is the measurement error. We make the assumptions that $\epsilon_{ji}(v_{jis})|\mathbf{b}_{ji}\sim N(0,\sigma_j^2) \text{ for } s=1,\dots,m_{ji}$ and $\epsilon_{ji}(v)$ and $\epsilon_{ji}(v')$ are independent for $v\neq v'.$ 

For the survival model, let $h_{ji}(t)$ be the hazard function for patient $i$ in subgroup $j$ and $h_{oj}$ be the baseline hazard function for subgroup $j$. Then the survival submodel takes the form
\begin{equation}
\label{eq:surv}
h_{ji}(t)=h_{0j}(t)\exp\{\gamma_j X_{ji}(t)+\theta_j \psi_{ji}\} \hspace{1cm} \text{for }j=1,2.
\end{equation}

\subsection{5-year restricted mean survival time (RMST)}
\label{subsec:RMST}
For the joint model of Equations~\eqref{eq:joint}---\eqref{eq:surv}, the coefficients $b_{2j}$ and $\theta_j$ are two treatment parameters where $\theta_j$ is the direct effect of treatment acting on survival and $b_{2j}$ is the indirect effect. We require a single one dimensional test statistic that summarises the overall effect of treatment and we propose using the restricted mean survival time (RMST) to do so. 

\citet{royston2011use} define RMST as the area under the survival curve up to time $t^*$. The value of $t^*$ is fixed at the design stage and we shall discuss our choice. Let $\bs{\eta}_j$ be the $p\times 1$ vector of all parameters in the joint model in subgroup $j$. Suppose that $F_{0j}$ and $F_{1j}$ are time-to-failure random variables for patients on the control and experimental treatment arms respectively and that $S_{0j}(t;\bs{\eta}_j)$ and $S_{1j}(t;\bs{\eta}_j)$ are the corresponding survival functions integrated over any patient specific random effects. Then the difference in RMST between treatment groups is
\begin{equation}
\label{eq:RMST}
\Delta_j(t^*;\bs{\eta}_j) = \mathbb{E}[\min(F_{1j},t^*)]-\mathbb{E}[\min(F_{0j},t^*)]= \int_0^{t^*} \left[S_{1j}(t;\bs{\eta}_j)-S_{0j}(t;\bs{\eta}_j)\right]dt.
\end{equation}

Most commonly, non-parametric methods are employed for estimating $\Delta_j(t^*;\bs{\eta}_j)$ when the proportional hazards assumption does not hold since the estimator is robust to model misspecification. Our motivation for using RMST is to find a test statistic summarising the effect of two treatment parameters. Hence we shall focus on the parametric RMST estimate which is now described.

Let the cumulative hazard function for patient $i$ in subgroup $j$ be $H_{ji}(t;\bs{\eta}_j)$. The control and experimental treatment survival functions integrated over the random effects are 
\begin{align*}
S_{0j}(t;\bs{\eta}_j) &= \int_{-\infty}^{\infty} \exp\{-H_{ji}(t;\bs{\eta}_j, \psi_{ji}=0\} f(b_{0ji},b_{1ji}) db_{0ji} db_{1ji}\\
S_{1j}(t;\bs{\eta}_j) &= \int_{-\infty}^{\infty} \exp\{-H_{ji}(t;\bs{\eta}_j, \psi_{ji}=1\} f(b_{0ji},b_{1ji}) db_{0ji} db_{1ji}
\end{align*}
where $ f(b_{0ji},b_{1ji})$ is the probability density function for the normal distribution of the random effects. Gauss-Hermite integration can be used to efficiently calculate the integrals over $b_{0ji}$ and $b_{1ji}$. The survival functions are then substituted into Equation~\eqref{eq:RMST} to calculate $\Delta_j(t^*;\bs{\eta}_j).$ \citet{rizopoulos2012joint} presents the full likelihood function for the joint model from which we can obtain the MLE $\hat{\bs{\eta}}^{(k)}_j$ at analysis $k.$ An estimate for the treatment difference is given by $\Delta_j(t^*;\hat{\bs{\eta}}_j^{(k)})$ for $k=1,\dots,K.$

The delta method by \citet{doob1935limiting} is used calculate the variance of the parametric RMST estimate. We have that $\hat{\bs{\eta}}^{(k)}_j$ has the same dimensionality as $\bs{\eta}_j$, a $p\times 1$ vector. Let $\Sigma_j^{(k)}$ be the $p\times p$ covariance matrix of the MLE $\hat{\bs{\eta}}_j^{(k)}$ at analysis $k$, then we have that $n^{1/2}(\hat{\bs{\eta}}_j^{(k)}-\bs{\eta}_j)\xrightarrow{d}N(\mathbf{0}, \Sigma_j^{(k)})$. The information level at analysis $k$ for the difference in RMST between treatment arms is given by
$$
\mathcal{I}^{(k)}_j = n_j\left(\left[\partial \Delta_j(t^*;\hat{\bs{\eta}}_j^{(k)})/\partial \bs{\eta}_j\right]^T\Sigma_j^{(k)} \left[\partial \Delta_j(t^*;\hat{\bs{\eta}}_j^{(k)})/\partial \bs{\eta}_j\right]\right)^{-1}
$$
where $\partial \Delta_j(t^*;\hat{\bs{\eta}}_j^{(k)})/\partial \bs{\eta}_j$ is the $p\times 1 $ vector which is the first derivative of the function $\Delta_j(t^*;\bs{\eta}_j)$ with respect to the vector $\bs{\eta}_j$ evaluated at $\hat{\bs{\eta}}_j^{(k)}$. In the calculation of $\mathcal{I}_j^{(k)},$ a consistent estimate $\hat{\Sigma}_j^{(k)}$ can be substituted in place of the covariance matrix $\Sigma_j^{(k)}$. In practice, the MLE $\hat{\bs{\eta}}^{(k)}_j$ and an estimate of the covariance matrix can be calculated using the R package JM by \citet{rizopoulos2012joint}.

The value of the truncation time $t^*$ is important for defining the estimand of the clinical trial and we believe that $t^*$ should be chosen to be clinically meaningful. Contrastingly, Royston and Parmar\cite{royston2013restricted}  suggest taking $t^*$ as the value that minimises the expected sample size given the recruitment time and minimum follow-up time. For our simulation study in the main text, each trial has recruitment roughly 2 years and final analysis time at roughly 5 years. To ensure that the method is robust to model misspecifications, it is important to avoid extrapolation of the RMST estimate beyond analysis time. Hence, we suggest using RMST at time $t^*=5$ which is primarily chosen to be clinically meaningful and secondly meets the suggestions.

To summarise the overall effect of treatment on survival incorporating both treatment parameters $b_{2j}$ and $\theta_j,$ we test the null hypothesis 
$$
H_{0j}:\Delta_j(5;\bs{\eta}_j)\leq 0 \hspace{1cm} \text{for }j=1,2,F.
$$
To do so, we define RMST estimates $\Delta_j(5;\hat{\bs{\eta}}_j^{(k)}),$ information levels $\mathcal{I}_j^{(k)}$ and $Z$-statistics for each $k=1,\dots,K$ and $j = 1,2,F.$

\subsection{Discussion of the use of RMST for enrichment trials}

There are benefits to using the RMST methodology to incorporate a causal effect ef treatment acting through the biomarker. However, this method is accompanied by additional challenges.

The RMST methodology suffers from the complication relating to the choice of the truncation time $t^*$. The choice of $t^*$ is not always well described and small differences can result in large changes in the operating characteristics of a trial. Therefore, $t^*$ requires careful consideration and planning.
The RMST methodology relies on finding maximum likelihood estimates and the model is overparameterised when any parameter is equal to zero. In such a case, the resulting covariance matrix is often not positive semi-definite and the analysis cannot be performed.

The advantage of the conditional score method is that no assumptions are required for either the distribution of the random effects or the functional form of the baseline hazard function. The RMST method does allow us to make inferences which leverage information about the treatment effect in the longitudinal data, however we have found that in comparison to the conditional score method, there is not much to be gained by doing so. 
\section{Joint density function in the full population}
\label{appendix:fZ}
To compute the joint density $f_{Z^{(1)}_W, W}(z^{(1)}_F, F;\bs{\Theta})$ we shall decompose $Z_F^{(1)}$ into the sum of two independent normal random variables and apply the convolution formula for probability density functions. Let $Z_F^{(1)} = X_1 + X_2$ where $X_1$ and $X_2$ are normally distributed random variables given by
\begin{align*}
X_1 &=  \lambda\sqrt{\mathcal{I}_F^{(1)}}\hat{\theta}_1^{(1)}\sim N\left( \lambda\sqrt{\mathcal{I}_F^{(1)}}\mu_1, \lambda^2\mathcal{I}^{(1)}_F/\mathcal{I}^{(1)}_1\right) \\
X_2 &=  \lambda\sqrt{\mathcal{I}_F^{(1)}}\hat{\theta}_2^{(1)}\sim N\left( \lambda\sqrt{\mathcal{I}_F^{(1)}}\mu_2, \lambda^2\mathcal{I}^{(1)}_F/\mathcal{I}^{(1)}_2\right).
\end{align*}
The threshold selection criterion is transformed so that the constraint $Z_1 > \zeta$ implies that $X_1 >\lambda\sqrt{\mathcal{I}_F^{(1)}}\zeta/\sqrt{\mathcal{I}_1^{(1)}}$ and the case $Z_2 > \zeta$ is transformed to $X_2 >\lambda\sqrt{\mathcal{I}_F^{(1)}}\zeta/\sqrt{\mathcal{I}_2^{(1)}}$. By the convolution formula for probability density functions, we have
\begin{align*}
&f_{Z^{(1)}_W|W}(z^{(1)}_F|W=F;\bs{\Theta}) = \int_{-\infty}^\infty f_{X_1|W}(u|W=1;\bs{\Theta})f_{X_2|W}(x_2-u|W=1;\bs{\Theta})du\\
&=\int_{-\infty}^\infty \frac{\sqrt{\mathcal{I}_1^{(1)}\mathcal{I}_2^{(1)}}  \phi\left(\frac{\sqrt{\mathcal{I}_1^{(1)}}(u-\lambda\sqrt{\mathcal{I}_F^{(1)}})}{\lambda\sqrt{\mathcal{I}_F^{(1)}}}\right)\phi\left(\frac{\sqrt{\mathcal{I}_2^{(1)}}(z_F^{(1)}-u-(1-\lambda)\sqrt{\mathcal{I}_F^{(1)}})}{(1-\lambda)\sqrt{\mathcal{I}_F^{(1)}}}\right)}{\lambda(1-\lambda)\mathcal{I}_F^{(1)}\left[1- \Phi\left(\zeta-\mu_1\sqrt{\mathcal{I}_1^{(1)}}\right)\right]\left[1- \Phi\left(\zeta-\mu_2\sqrt{\mathcal{I}_2^{(1)}}\right)\right]} du
\end{align*}

In the main paper, we defined $f_{Z^{(1)}_W, W}\left(z^{(1)}_w, w;\bs{\Theta}\right)= \mathbb{P}\left(W=w;\bs{\Theta}\right)f_{Z^{(1)}_W|W}\left(z^{(1)}_w|W=w;\bs{\Theta}\right)$. Therefore, combining this with the threshold selection criteria, we have
$$
f_{Z^{(1)}_W, W}(z^{(1)}_F, F;\bs{\Theta}) = \frac{\sqrt{\mathcal{I}_1^{(1)}\mathcal{I}_2^{(1)}}}{\lambda(1-\lambda)\mathcal{I}_F^{(1)}} \int_{-\infty}^\infty \phi\left(\frac{\sqrt{\mathcal{I}_1^{(1)}}(u-\lambda\sqrt{\mathcal{I}_F^{(1)}})}{\lambda\sqrt{\mathcal{I}_F^{(1)}}}\right)
\phi\left(\frac{\sqrt{\mathcal{I}_2^{(1)}}(z_F^{(1)}-u-(1-\lambda)\sqrt{\mathcal{I}_F^{(1)}})}{(1-\lambda)\sqrt{\mathcal{I}_F^{(1)}}}\right) du.
$$
\section{Proof of strong control of the FWER}
We now show that the threshold selection rule combined with an error spending test controls the FWER in the strong sense. We shall use the following definitions
\begin{align*}
S_1(\theta_1, \theta_2)&=\left\{Z_1^{(1)} > \zeta - \theta_1\sqrt{\mathcal{I}_1^{(1)}}\cap Z_2^{(1)} \leq \zeta - \theta_2\sqrt{\mathcal{I}_2^{(1)}}\right\} \\
S_2(\theta_1, \theta_2)&=\left\{Z_1^{(1)} \leq \zeta - \theta_1\sqrt{\mathcal{I}_1^{(1)}}\cap  Z_2^{(1)} > \zeta - \theta_2\sqrt{\mathcal{I}_2^{(1)}}\right\} \\
S_F(\theta_1, \theta_2)&=\left\{Z_1^{(1)} > \zeta - \theta_1\sqrt{\mathcal{I}_1^{(1)}}\cap Z_2^{(1)} > \zeta - \theta_2\sqrt{\mathcal{I}_2^{(1)}}\right\} \\
A_j^{(k)}(\theta_j) & = \left\{a_k - \theta_j\sqrt{\mathcal{I}_j^{(k)}} < Z_j^{(k)} < b_k - \theta_j\sqrt{\mathcal{I}_j^{(k)}}\right\} \\
B_j^{(k)}(\theta_j) & = \left\{Z_j^{(k)} >b_k - \theta_j\sqrt{\mathcal{I}_j^{(k)}}\right\}.
\end{align*}
The sets \(S_1(\theta_1,\theta_2),S_2(\theta_1,\theta_2),S_F(\theta_1,\theta_2)\) describe the selection criteria and the sets \(A_j^{(k)}(\theta_j)\) and \(B_j^{(k)}(\theta_j)\) represent the accept and reject regions of the hypothesis test \(H_j\) for \(j=1,2,F.\)

Suppose that \(\mathcal{L}=\{j=1,2,F|\theta_j\leq0\}\) is the set of indices corresponding to \emph{true} null hypotheses \(H_{0,j}, (j=1,2,F)\). Let \(R_j(\theta_j)\) be the event that \(H_{0j}\) is rejected and let \(\bar{R}(\theta_1,\theta_2)\) be the event that at least one true \(H_{0,j}\) is rejected. These are given by
\begin{align*}
R_j(\theta_j)&=\bigcup \limits_{k=1}^{K}\left[\left\{\bigcap\limits_{m=1}^{k-1}A_j^{(m)}(\theta_j)\right\}\cap B_j^{(k)}(\theta_j)\right] \\
\bar{R}(\theta_1,\theta_2)&=\bigcup\limits_{j\in\mathcal{L}} \left(S_j(\theta_1,\theta_2)\cap R_j(\theta_j)\right).
\end{align*}
In what follows, we aim to show that \(\bar{R}(\theta_1,\theta_2)\subseteq \bar{R}(0,0),\) which leads to showing that FWER is maximized under the global null. To do so, we impose the following condition.
\begin{conditions}\label{cond:strong_control}
The treatment effect in the full population, \(\theta_F=\lambda \theta_1+(1-\lambda)\theta_2,\) is non-negative.
\end{conditions}
We note here the similarity between Condition~\ref{cond:strong_control} and the condition in the proof by \citet{magnusson2013group}, where the authors make the assumption that treatment effects \(\theta_j\) cannot be negative for any \(j\). \citet{magnusson2013group} argue that treatment effects of opposite sign are ``rare and highly implausible". Our condition is not as restrictive, since treatment effects other than \(\theta_F\) are allowed to be negative. This condition ensures that the subgroup selected does not differ under scenarios \((\theta_1,\theta_2)\) and \((0,0)\). We believe that, with moderate information levels, these unaccounted events are so unlikely that they will not affect the FWER. Without this assumption however, it is possible to show that asymptotically the FWER is protected in the strong sense. As \(n\) increases, the information levels increase and it can be seen that \(\limsup\limits_{n\rightarrow\infty}S_i(\theta_1,\theta_2)\cap S_j(0,0) =\emptyset\) for any \(i\neq j\). Under Condition~\ref{cond:strong_control}, there are four possibilities for the configuration of \(\bs{\Theta}=(\theta_1,\theta_2)\) and these are 
\begin{enumerate}
\item \(\theta_1=\theta_2= \theta_F= 0\)
\item \(\theta_1 \leq 0, \theta_2> 0, \theta_F > 0\)
\item \(\theta_1 > 0, \theta_2 \leq 0, \theta_F > 0\)
\item \(\theta_1 > 0, \theta_2> 0, \theta_F > 0\)
\end{enumerate}

We are now equipped to prove that the threshold selection rule combined with an error spending test controls the FWER in the strong sense.
\begin{theorem}
\label{theorem:strong_control}
For global null hypothesis \(H_G\) and any \(\bs{\Theta}=(\theta_1,\theta_2)\) such that Condition~\ref{cond:strong_control} holds, we have
\[ \mathbb{P}(\text{Reject at least one true }H_j|\bs{\Theta})\leq \mathbb{P}(\text{reject at least one }H_j|H_G).\]
\end{theorem}
\begin{proof}

For this proof, we first show that \(\bar{R}(\theta_1,\theta_2)\subseteq\bar{R}(0,0)\) for each of the four cases which were possible under Condition~\ref{cond:strong_control}. For the first case, we have that \(\theta_1=\theta_2=\theta_F=0\) which is equivalent to the global null and we have \(\bar{R}(\theta_1,\theta_2)=\bar{R}(0,0).\)

For case 2, when \(\theta_1\leq 0, \theta_2 > 0, \theta_F >0\), the event that at least one true \(H_{0,j}\) is rejected is \(\bar{R}(\theta_1,\theta_2)=S_1(\theta_1,\theta_2)\cap R_1(\theta_1)\) and hence we show that \(S_1(\theta_1,\theta_2)\cap R_1(\theta_1)\subseteq S_1(0,0)\cap R_1(0)\). Suppose that \(x= (Z_1^{(1)},\dots,Z_1^{(K)},Z_2^{(1)},\dots,Z_2^{(K)})\in S_1(\theta_1,\theta_2)\cap R_1(\theta_1),\) so that \(Z_1^{(1)} > \zeta - \theta_1\sqrt{\mathcal{I}_1^{(1)}}\) and \(Z_2^{(1)} \leq \zeta-\theta_2\sqrt{\mathcal{I}_2^{(1)}}\). But \(\theta_1 \leq 0 \) and \(\theta_2>0\) so \(Z_1^{(1)} > \zeta\) and \(Z_2^{(1)} \leq \zeta\) so \(x\in S_1(0,0)\). Following the work of \citet{magirr2012generalized}, we also have that \(x\in \bigcup\limits_{k=1}^{K}\left[\left\{\bigcap\limits_{m=1}^{k-1}A_1^{(m)}(\theta_1)\right\}\cap B_1^{(k)}(\theta_1)\right]\). For some \(k\in\{1,\dots,K\},\) \(Z_1^{(k)}\in B_1^{(k)}(\theta_1)\) and \(Z_1^{(m)}\in A_1^{(m)}(\theta_1)\) for \(m=1,\dots,k-1\). \(Z_1^{(k)}\in B_1^{(k)}(\theta_1)\) implies that \(Z_1^{(k)}\in B_1^{(k)}(0)\) and \(Z_1^{(m)}\in A_1^{(m)}(\theta_1)\) implies that \(Z_1^{(m)}\in A_1^{(m)}(0)\cup B_1^{(m)}(0)\) for \(m=1,\dots,k-1.\) Therefore \(x\in \bigcup\limits_{k=1}^{K}\left[\left\{\bigcap\limits_{m=1}^{k-1}A_1^{(m)}(0)\right\}\cap B_1^{(k)}(0)\right]\). Hence, we have the result  \(\bar{R}(\theta_1,\theta_2)=\bar{R}(0,0).\)

Case 3 can be shown by exactly the same argument as for case 2, replacing all indices \(j=1\) with \(j=2\). Finally, Case 4 is trivial since this is the case where none of the hypotheses \(H_{0j}, (j=1,2,F)\) are true. Hence the event that at least one true \(H_{0,j}\) is rejected is \(\bar{R}(\theta_1,\theta_2)=\emptyset.\) In each of the four cases we have \(\bar{R}(\theta_1,\theta_2)\subseteq\bar{R}(0,0)\) and therefore
\begin{align*}
\mathbb{P}\{\text{Reject at least one true }H_j|\theta_1,\theta_2\} &= \mathbb{P}\{\bar{R}(\theta_1,\theta_2)\}\\
&\leq \mathbb{P}\{\bar{R}(0,0)\}\\
& =\mathbb{P}\{\text{Reject at least one true }H_j|H_0\}.
\end{align*}
\end{proof}
\section{Additional results from simulation studies}
We present the results from the simulation study in the results section of the main paper. To recap, for one simulation; generate a dataset of patients from the joint model, then subgroup selection and decisions about $H_0$ are performed after $d_1^{(1)}$ and $d^{(2)}$ events have been observed. During each simulation run, all four methods are evaluated using the same dataset and after the same number of events. This is so that differences in the trial results can be attributed to the analysis methodology and not trial design features. The simulations are repeated $N=10^4$ times for each set of parameter values.

FWER for the simulation studies are shown in Web Table~\ref{tbl:type1}. Results confirm that the FWER are close to 0.025 for $N=10^4$ simulations.
\begin{table}
\centering
\caption{Family wise error rates (FWER) for each method.\label{tbl:type1}}
\begin{tabular}{|lll|ccc|}
\hline
&&&\multicolumn{3}{c}{FWER}\\
$\bs{\gamma}$ & $\bs{\sigma}^2$ & $\bs{\phi}_2$ &
Conditional score & Cox & Cox with biomarker
\\\hline
\rule{0pt}{3.5ex}$\mathbf{0}$ & 
     $\mathbf{0.25}$ &
     $\mathbf{5}$ &0.026 &0.024 &0.025 \\
$\mathbf{0.4}$ & 
     $\mathbf{0.25}$ &
     $\mathbf{5}$ &0.025 &0.026 &0.026 \\
$\mathbf{0.8}$ & 
     $\mathbf{0.25}$ &
     $\mathbf{5}$ &0.024 &0.024 &0.019 \\
$\mathbf{1.2}$ & 
     $\mathbf{0.25}$ &
     $\mathbf{5}$ &0.026 &0.018 &0.014 \\
\rule{0pt}{3.5ex}$\mathbf{0.8}$ & 
     $\mathbf{0}$ &
     $\mathbf{5}$ &0.028 &0.025 &0.024 \\
$\mathbf{0.8}$ & 
     $\mathbf{0.25}$ &
     $\mathbf{5}$ &0.024 &0.024 &0.019 \\
$\mathbf{0.8}$ & 
     $\mathbf{1}$ &
     $\mathbf{5}$ &0.020 &0.023 &0.011 \\
$\mathbf{0.8}$ & 
     $\mathbf{2.25}$ &
     $\mathbf{5}$ &0.025 &0.029 &0.014 \\
\rule{0pt}{3.5ex}$\mathbf{0.8}$ & 
     $\mathbf{0.25}$ &
     $\mathbf{0}$ &0.026 &0.025 &0.018 \\
$\mathbf{0.8}$ & 
     $\mathbf{0.25}$ &
     $\mathbf{2.5}$ &0.021 &0.022 &0.016 \\
$\mathbf{0.8}$ & 
     $\mathbf{0.25}$ &
     $\mathbf{5}$ &0.024 &0.024 &0.019 \\
$\mathbf{0.8}$ & 
     $\mathbf{0.25}$ &
     $\mathbf{7.5}$ &0.026 &0.025 &0.019 \\
\hline
\end{tabular}
\end{table}

Selection probabilities, denoted $\mathbb{P}(\text{Select }S_1)$ in Web Table~\ref{tbl:results}, are calculated as the proportion of simulations which select subgroup $S_1.$ This is an appropriate summary metric since $S_1$ is the subgroup which truly benefits from the experimental treatment. The value of $d^{(1)}_1$ has been calculated to ensure $\mathbb{P}(\text{Select }S_1)=0.6$ using the conditional score method. Web Table~\ref{tbl:results} confirms that the probabilities are close to 0.6 for $N=10^4$ simulations. The value of $d^{(2)}$, has been calculated with reasonable accuracy since power is suitably close to 0.9 for $N=10^4$ in all cases for the conditional score method. The alternative methods do not attain the desired selection probabilities because the trial is designed using the conditional score analysis method.  Generally, the Cox method has lower selection probabilities than the conditional score method and the Cox with biomarker has higher selection probabilities than the Conditional score method. These results become more extreme for increases in $\gamma.$

\begin{table}
\centering
\caption{Selection probabilities and power for each method.\label{tbl:results}}
\begin{tabular}{|lll|ccc|ccc|}
\hline
&&&\multicolumn{3}{c}{$\mathbb{P}(\text{select }S_1)$}&
\multicolumn{3}{c}{Power}\\
&&&Conditional & Cox & Cox with &
Conditional & Cox & Cox with \\
$\bs{\gamma}$ & $\bs{\sigma}^2$ & $\bs{\phi}_2$ &
score &&biomarker &score&&biomarker
\\\hline
\rule{0pt}{3.5ex}$\mathbf{0}$ & 
     $\mathbf{0.25}$ &
     $\mathbf{5}$ &0.588 &0.610 &0.604 &0.908 &0.915 &0.914 \\
$\mathbf{0.4}$ & 
     $\mathbf{0.25}$ &
     $\mathbf{5}$ &0.636 &0.541 &0.626 &0.909 &0.616 &0.897 \\
$\mathbf{0.8}$ & 
     $\mathbf{0.25}$ &
     $\mathbf{5}$ &0.625 &0.431 &0.621 &0.907 &0.570 &0.887 \\
$\mathbf{1.2}$ & 
     $\mathbf{0.25}$ &
     $\mathbf{5}$ &0.611 &0.363 &0.618 &0.903 &0.440 &0.862 \\
\rule{0pt}{3.5ex}$\mathbf{0.8}$ & 
     $\mathbf{0}$ &
     $\mathbf{5}$ &0.628 &0.424 &0.619 &0.904 &0.560 &0.897 \\
$\mathbf{0.8}$ & 
     $\mathbf{0.25}$ &
     $\mathbf{5}$ &0.625 &0.431 &0.621 &0.907 &0.570 &0.887 \\
$\mathbf{0.8}$ & 
     $\mathbf{1}$ &
     $\mathbf{5}$ &0.637 &0.438 &0.627 &0.905 &0.590 &0.851 \\
$\mathbf{0.8}$ & 
     $\mathbf{2.25}$ &
     $\mathbf{5}$ &0.631 &0.458 &0.631 &0.906 &0.610 &0.787 \\
\rule{0pt}{3.5ex}$\mathbf{0.8}$ & 
     $\mathbf{0.25}$ &
     $\mathbf{0}$ &0.628 &0.503 &0.617 &0.910 &0.780 &0.915 \\
$\mathbf{0.8}$ & 
     $\mathbf{0.25}$ &
     $\mathbf{2.5}$ &0.612 &0.433 &0.600 &0.907 &0.620 &0.883 \\
$\mathbf{0.8}$ & 
     $\mathbf{0.25}$ &
     $\mathbf{5}$ &0.625 &0.431 &0.621 &0.907 &0.570 &0.887 \\
$\mathbf{0.8}$ & 
     $\mathbf{0.25}$ &
     $\mathbf{7.5}$ &0.643 &0.420 &0.627 &0.909 &0.480 &0.858 \\
\hline
\end{tabular}
\end{table}

We also present the expected number of hospital visits per patient and the expected stopping time for each method. In Web Table~\ref{tbl:extras}, the expected number of hospital visits is calculated as the mean number of longitudinal observations across all patients enrolled in the study across all simulations and the expected stopping time is the average study duration in years. It is challenging to make comparisons between methods for these outcomes. This is because the number of patients is not the same between methods due to capability of each method to select the correct subgroup. For example, it would appear at first sight that the simple Cox model method is most efficient because it results in the shortest average stopping time, for all cases. However, we find that this method has a relatively low chance of selecting any subgroup and often the trial stops at the first interim analysis to declare that the treatment is inefficacious in all subgroups. We have not included the number of hospital visits per patient for the simple Cox model and this is to highlight an advantage of the method that we do not need to collect the longitudinal data at all in this case. It is clear, however, that the trends in number of hospital visits and stopping times all follow the same structure as the trends in the probability of selecting subgroup $S_1$. Hence, for large $\gamma,\phi_2$ and $\sigma$, we require patients to have more regular blood tests and also the trial will take longer to reach a decision.
\begin{table}
\centering
\caption{Expected number of hospital visits per patien and expected stopping times
for each method.\label{tbl:extras}}
\begin{tabular}{|lll|ccc|ccc|}
\hline
&&&\multicolumn{3}{c}{$\mathbb{E}(\text{number of hospital visits per patient})$}&
\multicolumn{3}{c}{$\mathbb{E}(\text{stopping time})$}\\
&&&Conditional & Cox & Cox with &
Conditional & Cox & Cox with \\
$\bs{\gamma}$ & $\bs{\sigma}^2$ & $\bs{\phi}_2$ &
score &&biomarker &score&&biomarker
\\\hline
\rule{0pt}{3.5ex}$\mathbf{0}$ & 
     $\mathbf{0.25}$ &
     $\mathbf{5}$ &11.1 &-&11.3 &2.96 &2.98 &2.98 \\
$\mathbf{0.4}$ & 
     $\mathbf{0.25}$ &
     $\mathbf{5}$ &12.0 &-&11.9 &3.32 &3.22 &3.31 \\
$\mathbf{0.8}$ & 
     $\mathbf{0.25}$ &
     $\mathbf{5}$ &11.9 &-&11.9 &3.41 &3.11 &3.42 \\
$\mathbf{1.2}$ & 
     $\mathbf{0.25}$ &
     $\mathbf{5}$ &11.7 &-&11.7 &3.57 &3.13 &3.57 \\
\rule{0pt}{3.5ex}$\mathbf{0.8}$ & 
     $\mathbf{0}$ &
     $\mathbf{5}$ &11.8 &-&11.8 &3.26 &3.00 &3.30 \\
$\mathbf{0.8}$ & 
     $\mathbf{0.25}$ &
     $\mathbf{5}$ &11.9 &-&11.9 &3.41 &3.11 &3.42 \\
$\mathbf{0.8}$ & 
     $\mathbf{1}$ &
     $\mathbf{5}$ &12.4 &-&12.0 &3.96 &3.52 &3.87 \\
$\mathbf{0.8}$ & 
     $\mathbf{2.25}$ &
     $\mathbf{5}$ &12.5 &-&12.1 &4.56 &3.86 &4.27 \\
\rule{0pt}{3.5ex}$\mathbf{0.8}$ & 
     $\mathbf{0.25}$ &
     $\mathbf{0}$ &11.3 &-&11.2 &3.07 &2.93 &3.05 \\
$\mathbf{0.8}$ & 
     $\mathbf{0.25}$ &
     $\mathbf{2.5}$ &11.6 &-&11.5 &3.23 &2.98 &3.23 \\
$\mathbf{0.8}$ & 
     $\mathbf{0.25}$ &
     $\mathbf{5}$ &11.9 &-&11.9 &3.41 &3.11 &3.42 \\
$\mathbf{0.8}$ & 
     $\mathbf{0.25}$ &
     $\mathbf{7.5}$ &12.1 &-&11.9 &3.47 &3.11 &3.46 \\
\hline
\end{tabular}
\end{table}

\section{Details of the estimation of treatment effect estimates in alternative Cox proportional hazards models}
\subsection{Cox proportional hazards model}
\label{subsec:cox}
We now give an overview of how treatment effect estimates and information levels are calculated for each of the alternative Cox proportional hazards models. As a reminder, we introduce the model again. Denote $h_{0j}(t)$ as the baseline hazard function, $\theta_j$ the treatment effect and $\psi_{ji}$ as the treatment indicator that patient $i$ in subgroup $j=1,2$  receives the new treatment. Then the hazard function for the survival model is given by
\begin{equation}
\label{eq:cox_model}
h_{ji}(t)=h_{0j}(t)\exp\{\theta_j \psi_{ji}\}.
\end{equation}
Let $t_{ji}^{(k)}$ be the observed event time and let $\delta_{ji}^{(k)}$ be the observed censoring indicator for patient $i$ in subgroup $j=1,2$ at analysis $k$. Then $Y_{ji}^{(k)}(t)=\mathbb{I}\{t_{ji}^{(k)} \geq t\}$ is the at-risk process and $dN_{ji}^{(k)}(t) = \mathbb{I}\{t \leq t_{ji}^{(k)} < t+dt, \delta_{ji}^{(k)}=1\}$ is the counting process. As in \citet{jennison1997group}, the function $E_j^{(k)}(u, \theta_j)$ and the score function $U_j^{(k)}(\theta_j)$ at analysis $k$ are given by
\begin{equation}
\label{eq:score_2}
\begin{split}
E_j^{(k)}(t,\theta_j) &= \frac{\sum_{i=1}^{n_j} \psi_{ji}\exp\{\theta_j\psi_{ji}\}Y_{ji}^{(k)}(t)}{\sum_{i=1}^{n_j} \exp\{ \theta_j \psi_{ji}\}Y_{ji}^{(k)}(t)} \\
U_j^{(k)}(\theta_j) &= \int_0^{\tau_k}\sum_{i=1}^{n_j}\left(\psi_{ji}- E_j^{(k)}(t,\theta_j)\right) d\tilde{N}_{ji}^{(k)}(t).
\end{split}
\end{equation}

The function $U_j^{(k)}(\cdot)$ is a score function and $\theta_j$ can be estimated by solving the equation $U_j^{(k)}(\theta_j)=0$ for $\theta_j$. Let this estimate, at analysis $k$, be denoted by $\hat{\theta}_j^{(k)}$. By standard results by \citet{jennison1997group}, the estimates $\hat{\theta}_j^{(k)}$ follow the CJD where the information level at analysis $k$ is given by $\mathcal{I}^{(k)}_j = n_j\left[ \partial U^{(k)}_j(\hat{\theta}_j^{(k)})/\partial \theta_j\right]^{-1}.$

\subsection{Cox proportional hazards model with longitudinal data as a time-varying covariate}
\label{subsec:cox_time_varying}
As a reminder, $W_{ji}(v_{ji1}),\dots,W_{ji}(v_{jim_{ji}})$ are the observed values of the biomarker for patient $i$ in subgroup $j$ at times $v_{j-1},\dots, v_{jim_{ji}}$ and the definitions of the the at-risk process $Y_{ji}^{(k)}(t)$ and counting process function $dN_{ji}^{(k)}(u)$ are as in Section~\ref{subsec:cox}. Let $\gamma_j$ and $\theta_j$ be longitudinal data and treatment coefficients respectively, then the hazard function is given by
\begin{equation}
\label{eq:cox_long}
h_{ji}(t)=h_{j0}(t)\exp\{\gamma_jW_{ji}(t) + \theta_j \psi_{ji}\}.
\end{equation}
The function $E^{(k)}_j(u,\cdot)$ and score statistic $U^{(k)}_j(\cdot) $ for this model are given by
\begin{align}
\notag
E^{(k)}_j(t,\theta_j) &= \frac{\sum_{i=1}^{n_j} \{W_{ji}(t), \psi_{ji}\}^T\exp\{\gamma_jW_{ji}(t) + \theta_j \psi_{ji}\}Y_{ji}^{(k)}(t)}{\sum_{i=1}^{n_j} \exp\{\gamma_jW_{ji}(t) + \theta_j \psi_{ji}\}Y_{ji}^{(k)}(t)} \\
\label{eq:score_3}
U^{(k)}_j(\theta_j) &= \int_0^{\tau_k}\sum_{i=1}^{n_j}\left(\{W_{ji}(t), \psi_{ji}\}^T- E^{(k)}_j(t,\theta_j)\right) dN_{ji}^{(k)}(t).
\end{align}
Both objects $E^{(k)}_j(u,\theta_j)$ and $U^{(k)}_j(\theta_j) $ are $2\times 1$ dimensional vectors. To evaluate these objects, we will need to know $W_{ji}(t_{js})$ which is the value of the time-varying covariate for patient $i$ in subgroup $j$, evaluated at the event time of patient $s$ in subgroup $j$. For this model, $W_{ji}(\cdot)$ is a function of time and is known. For calculation purposes, for $t > v$, we shall set $W_{ji}(t)$ as $W_{ji}(v)$ where $v=max(v_{jim}|v_{jim}\leq t)$.

Again, the function $U_j^{(k)}(\cdot)$ is a score function and $\theta_j$ can be estimated by solving the equation $U_j^{(k)}(\theta_j)=0$ for $\theta_j$ and the information level at analysis $k$ can be calculated as $\mathcal{I}^{(k)}_j = n_j\left[ \partial U^{(k)}_j(\hat{\theta}_j^{(k)})/\partial \theta_j\right]^{-1}.$

\bibliographystyle{plainnat}  
\bibliography{enrichment}